\documentclass[letterpaper,10pt]{article}

\usepackage{amssymb}
\usepackage{amsmath}
\usepackage{psfig}

\def\nueo{\nu^{e_1}}
\def\nuet{\nu^{e_2}}

\def\nuret{\nu_R^{e_2}}
\def\nuleo{\nu_L^{e_1}}

\def\etr{e_{2R}}
\def\eol{e_{1L}}

\def\circle{\bigcirc}
\def\up{\overset{\uparrow}{{\circle}}}
\def\down{\underset{\downarrow}{{\circle}}}
\def\both{\overset{{\uparrow}}{\underset{{\downarrow}}{{\circle}}}}

\def\superspacyformat{
\setlength{\textheight}{8in}
\setlength{\textwidth}{5.7in}
\setlength{\evensidemargin}{0.2in}
\setlength{\oddsidemargin}{0.2in}
\setlength{\headheight}{0in}
\setlength{\headsep}{20pt}
\setlength{\topsep}{0in}
\setlength{\topmargin}{0.0in}
\setlength{\itemsep}{10pt}
\setlength{\parindent}{0pt}
\parskip=0.080in
}

\superspacyformat

\begin{document}

\title{..........................................................cobah-232-2000 \\
.\\
A Model of Elementary Particle Interactions}


\author{\it{Irshadullah Khan, Department of Physics} \\
\it{The College of Bahamas, Nassau, Bahamas}\\
email: irshadk@hotmail.com and ikhan@cob.edu.bs}
\maketitle

\begin{abstract}

There is a second kind of light which does not interact with our electrons.
However it interacts with some of our protons ($p$) and some of our neutrons ($n$) which are
both of two kinds $(\mbox{protons}\; :p,\; p' ,\; \mbox{neutrons} \;: n',\;n)$
differing in the two kinds of charges $(Q_1, Q_2)$ associated with the two kinds of light.
$p [p']$ and $n' [n]$ have $(Q_1, Q_2)$ values equal to (1,1) [(1,0)] and (0,0) [(0,1)] respectively.
There is also a second kind of electron $(Q_2 =1, Q_1= 0)$, equal in mass to our electron  $(Q_1 = -1, Q_2=0)$, ,
which does not interact with our kind of light. Three major scenarios  $S_1, \; S_2 \; and \;X_4$ arise.
In $S_1$, matter in the solar system on large scales is predominantly neutralized in both kinds  of charges 
and the weak forces of attraction among the sun and planets are due to a fundamental force of nature. However in
this scenario we must postulate that human consciousness is locked on to chemical reactions
in the retina involving the first kind of light and the first kind of electrons only. It is oblivious
to the simultaneous parallel chemical reactions governed by a chemistry  which is based on  
the second kind of light and the second kind of anti-electrons and involves the same physical atoms 
manifesting different atomic numbers $ Z' (= Q_2)$. In scenario $S_2$, matter in the solar system on large scales is
predominantly neutralized in the first kind of charge only. In this scenario human consciousness
is not restricted in its awareness to a narrowly ......
\end{abstract}

.\\
\\
\\
\\
\\
\\
\\
\\
\\
\\
\\
\\
\\
\\
\\
\\
\\
\\
\\
\\

Abstract continued........

circumscribed domain of physical reality.
However in this scenario we must accept the existence of very strong forces of attraction
in the solar system in order to counteract the strong repulsive forces due to the second
kind of charges on the sun and planets. The residual weak force of gravity can not be a fundamental
force. In this scenario the problem of unification is more tractable since all the fundamental forces are
comparable in strength. Furthermore this scenario allows dark matter to be made up of atoms having nuclei constituted 
of anti-baryons ($Q_1 < 0, Q_2 < 0 $) surrounded by 
shells of second kind of electrons. The shells of our electrons ($Q_1 < 0, Q_2 = 0 $) surrounding our nuclei 
($Q_1 > 0, Q_2 > 0 $) and the shells of second kind of electrons($Q_1=0, Q_2 > 0 $) surrounding the dark matter nuclei
($Q_1< 0, Q_2 < 0$) help prevent close encounters between
the baryons and anti-baryons trapped within their shells of different kinds. [If this scenario
holds then dark matter atoms may be harvested for a clean source of energy obtained in collisions with targets composed 
of our atomic nuclei ]. Scenario $X_4$ envisages the possibility of the two parallel chemistries being
identical and producing effects merely reinforcing each other and not causing matter to evolve
along competing tracks of chemical evolution. One symmetric  version of this scenario allows us to make
the following definite prediction: There is a neutron $(n')$ whose mass is equal to that of the proton $(p)$ and a proton $(p')$ whose mass is equal to that 
of the neutron $(n)$. $p'$ decays with a lifetime equal to the lifetime of the neutron ($n$). It decays according to
\begin{eqnarray*}
p' \rightarrow p + \overline{e_1} + \overline{{\nu}^{e_2}}&\hspace{0.3 in}& p' \rightarrow n' + \overline{e_1} + {\nu}^{e_1}\\
p' \rightarrow p + \overline{e_2} + \overline{{\nu}^{e_1}}&\hspace{0.3 in}& p' \rightarrow n' + \overline{e_2} + {\nu}^{e_2}
\end{eqnarray*}

Since the uncharged neutrino ${\nu}^{e_1}$ is almost undetectable and the charged anti-neutrino $\overline{{\nu}^{e_2}} $
events have not been recognized for what they are (viz., the misinterpreted jet events at high energies), the first two
of these decays of $p'$ mimic some aspects of the decays of  an anti-neutron and have probably been missed being noticed. In this scenario 
the observed part of the solar neutrino flux, being measured by experiments detecting 
${\nu}^{e_1}$, is  expected to be one fourth of the result derived 
from the standard model (and the solar model) if  the modifications introduced by the present model 
into the solar model itself are ignored. Another version of this scenario has $m(p')=m(p)$ and $m(n')=m(n)$ and is
ruled out by experiments. Implications of consequences of the model for the origin of cosmic microwave background, the nature of the Great
Attractor, masses of isomers of nuclear isotopes, separation of regions of spin and charge in high $T_c$ superconductors
and possible role for non-coding segments in DNA are briefly mentioned. Several other minor scenarios are briefly described. 
Experiments to decide between the various scenarios are proposed. 


\section{Questions}

We have been trying to make a physical theory that can provide us with
answers to the following questions:

\begin{enumerate}
\item
Why is there an apparent left and right as-symmetry in elementary
particle interactions? Is it really present at the presumably fundamental level?

\item
Is the appearance of matter and anti-matter as-symmetry real?  Is there a
more convincing explanation than the so-called Anthropic Principle \cite{Barrow} ,
whose proponents would probably suggest that the as-symmetry is real since
it seems indispensable for local survival of life?

Besides, the Anthropic principle is really an antithesis of the Cosmological
Principle (or the Copernican Doctrine) in disguise. Could we then avoid invoking 
contradictory principles to explain different aspects of the cosmos?

\item
Can the apparent solar neutrino deficit be explained?  Is there
a more natural explanation than to assume a mass for the neutrino,
in the absence of any direct experimental evidence for it.  Would it not be
preferable to avoid the element of arbitrariness thereby introduced in the
theory?

\item
What is the nature of the supernova \cite{Aschenbach} remnant which was
not observed as a supernova, even though it presumably has occurred around 500 AD, a
well-documented period of human history?

\item
What are the gamma ray bursts?  Is there a less contrived
explanation than colliding neutron stars \cite{Nishida}?

\item

Why is it that the smooth power-law cosmic ray energy spectrum does not fall off 
abruptly near $5 \;10^{19} eV$ ( the Greisen-Zatsepin-Kuzmin (GZK) threshold \cite{GKZ}, \cite{Zatsepin}), if the cosmic
background radiation (CBR)   \cite{Penzias} extends all around us?  If our estimates of the distances of 
the potential sources of very
high energy cosmic rays, that might have
generated  cosmic rays with energies above the GZK threshold, are correct then these
cosmic rays could not have survived the
inelastic interactions with the microwave photons, {\em if the} CBR {\em really exists all the way
to these sources of very high energy cosmic rays}.

Is there another explanation, perhaps less radical, than the recently proposed breakdown
of Lorentz invariance at very high energies \cite{Coleman}? 

\item
What is the nature of the CBR?  Is there an explanation that does not appeal to an initial singularity?
This has been a major outstanding issue since cosmological red-shift was first 
understood as a natural consequence of a principle more general than special relativity
(principle of reciprocity) \cite{Khan1, Khan7} in a non-expanding flat universe. This
effort has ( besides prompting investigations of solutions of non-linear problems \cite{Khan4}
involving functional iterates of order $\frac{1}{2}$) led to new ideas
about the possible fractal-like nature of time measurements \cite{Khan2, Khan3, Khan8}.
\\
Parenthetical remarks:
[A very brief summary of the idea for the proposed explanation of cosmological red-shift
given in \cite{Khan1} can be presented here. Consider the situation of the twins, carrying
clocks, who are together, get separated from each other for some time and then meet again, all the time
moving uniformly along straight lines except when at a certain instant their velocity relative to each other is
reversed. It is argued in \cite{Khan1} that, in spite of the fact that special relativity holds for uniform
relative motion, the two clocks initially in agreement also agree when the twins meet again; because each clock as seen
by the twin who is not carrying it performs a sudden forward jump in time at the moment of velocity
reversal. This forward jump is precisely equal to the time lag (of the other twin's clock) seen by the same twin during
the outward and inward uniform motions. Thus inward acceleration is seen to speed up a clock
at a rate which increases with acceleration as well as distance of the clock from the observer. In the 
case of radiating atoms in galaxies the acceleration due to galactic gravitational fields is outward relative 
to observers on earth. The radiating atoms are therefore seen to produce red-shifted light
since the clocks at the location of radiating atoms are being observed to be slowed down due to their outward acceleration
in galactic gravitational fields. Application of this argument gives a numerically correct relation between
the Hubble constant, the mass and linear size of
a typical galaxy, the velocity of light and Newton's constant of gravitation. This effect of acceleration
on clocks is quite distinct from the usual gravitational red-shift( in Einstein's theory of gravitation)-the latter
is derivable from the minimal form of the equivalence principle without invoking non-Euclidean geometry.

A more precise formulation of this idea, which yields the shapes of {\em apparent} rotation velocity
- central distance profiles for galaxies in agreement with observations,
has not yet been submitted for publication  because of the strong currents of opposition in the 
scientific establishment towards works with a potential for promoting 
unacceptable views about the cosmos. This hostility is manifested, for example, by the systematic obstructions which forced
the heroically courageous Halton Arp \cite{Arp}, \cite{Bertola}, who persisted in drawing attention to observational
evidence for instances of  gross non-variation of distance with  red-shift (GRINDERS), 
into reluctant retirement \cite{Hoyle}.]

\end{enumerate}

The above questions will all be addressed by a new model of electroweak
particle interactions which is presented in this paper.

\section{The Model}

\subsection{Symmetries}

Apart from the symmetries associated with the Poincar$\acute{e}$ group of
space-time transformations, the model has the symmetries of two
gauge groups of rotations in three dimensions; $R_1, R_2$ and two
$U(1)$ gauge groups of phase transformations; $U_1, U_2$.

Associated with $R_1, R_2$ are corresponding (Yang-Mills) gauge fields
$W_\mu^{1,j}, W_\mu^{2,j}$, $(j=1,2,3$; $\mu=0,1,2,3)$.
Infinitesimal elements in the Poincar$\acute{e}$ and gauge groups
correspondingly generate (inhomogeneous) linear actions with respect to
the indices $\mu$ and $j$, respectively. These actions are local.
We associate the same dimensionless coupling
constant $g$ with $W_\mu^{a,j}$ $(a=1,2)$.

The generators of infinitesimal transformations in $R_1,R_2$ are
denoted $T^j, S^j$ respectively and satisfy:
\begin{eqnarray*}
\left[ T^j, T^k \right] & =  & i \epsilon^{jkl} T^l\\
\left[ S^j, S^k \right] & =  & i \epsilon^{jkl} S^l\\
\left[ T^j, S^k \right] & = &  = 0
\end{eqnarray*}

Associated with $U_1, U_2$ are corresponding gauge fields $B^1_\mu$ and $B^2_\mu$.
The same dimensionless coupling constant, $f$, is associated with the $B^a_\mu \;$s.

The generators of the groups $U_1, U_2$ are designated $U^1, U^2$
respectively, and satisfy:
\begin{eqnarray*}
\left[ U^1, U^2 \right] & = & 0\\
\left[ U^a, T^j \right] & = & 0 \\
\left[ U^a, S^j \right] & = & 0
\end{eqnarray*} 

The generators of the gauge groups commute with the generators of the Poincar$\acute{e}$ group.
There is also a discrete symmetry transformation which is described in sub-section (\ref{parity}),
page (\pageref{parity}).

\subsection{Structure of the Model}
\label{sec2.2}
The model has two kinds of quarks and two kinds of leptons in each
family; all are four component Dirac spinors.  For example, in the
first family we have quarks; $u_1$, $d_1$ and $u_2$, $d_2$ and
leptons; $\nu^{e_1}$, $e_1$ and $\nu^{e_2}$, $e_2$.   The 
description that follows will be presented in terms of the first
family, though its content applies analogously to the other families.  
The leptons in the first family, that are subscripted by $1$, have the same quantum
numbers as the leptons of the standard model \cite{G1, Weinberg1, Salam, GIM}.

The model has two kinds of photons
\begin{eqnarray}
\label{A-mu-define}
\begin{array}{rcl}
{\cal A}_\mu^a & = & \frac{1}{\sqrt{f^2+g^2}}(f W_\mu^{a,3} + g B_\mu^a)\\
a & = & 1,2
\end{array}
\end{eqnarray}

and associated charges $Q_1$ and $Q_2$.  The assignment of these
charges and the baryonic number $B$ for quarks and leptons is given in Table (1).
\\
\\

\begin{center}
\label{tab1}
\begin{tabular}{|r||c|c|c|}
\hline
   & $Q_1$ & $Q_2$ & $B$\\
\hline
\hline
$u_1$   & $1/3$  & $1/3$ & $-1/3$ \\
$d_1$   & $-2/3$ & $1/3$ & $-1/3$ \\
$u_2$   & $2/3$  & $2/3$ & $1/3$  \\
$d_2$   & $-1/3$ & $2/3$ & $1/3$  \\
\hline
$\nueo$ & $0$    & $0$   & $0$    \\
$e_1$   & $-1$   & $0$   & $0$    \\
$\nuet$ & $1$    & $1$   & $0$    \\
$e_2$   & $0$    & $1$   & $0$    \\
\hline
\end{tabular}
\end{center}
\begin{center}
\begin{tabular}{c}
Table (1)
\end{tabular}
\end{center}
The spontaneous breakdown of symmetry generated by non-vanishing
vacuum expectation values of scalar Higgs fields \cite{Higgs1, Higgs2, Higgs3, Ander1, Brout} (details of this aspect 
of the model will not be presented here) yield four
charged massive vector bosons and two 
neutral massive vector bosons ${\cal Z}_\mu^1, {\cal Z}_\mu^2$. 
The four charged massive vector bosons $W_\mu^{a\pm} = \frac{1}{\surd{2}}(W_\mu^{a,1} \pm i W_\mu^{a,2})$ have equal
masses $M_W$, while the two neutral massive vector bosons ${\cal
Z}_\mu^a =\frac{1}{\sqrt{f^2+g^2}} (g W_\mu^{a,3} - f B_\mu^a)$ have equal masses $M_{\cal Z}$.
The two kinds of electrons, muons, $\tau$ leptons are assumed to have correspondingly equal masses -i.e. $m(e_1) =m(e_2)$. 
etc. For the two kinds of quarks the corresponding equality of masses $[(m(u_1) = m(u_2), m(d_1)= m(d_2)]$ holds in one 
of the scenarios considered.  Another scenario in which $m(u_1)=m(d_2)$ and $m(u_2)=m(d_1)$ is also considered. The $Q_1$
and $Q_2$ charges of the weak interaction bosons are given in Table (2).

\begin{center}
\begin{tabular}{|r||c|c|c|c|c|c|}
\hline
           & $W_\mu^{1+}$ & $W_\mu^{1-}$ & $W_\mu^{2+}$ & $W_\mu^{2-}$ & ${\cal Z}_\mu^1$ & ${\cal Z}_\mu^2$ \\
\hline
\hline
$Q_1$      & $-1$ & $+1$ & $0$ & $0$ & $0$ & $0$ \\
$Q_2$      & $0$ & $0$ & $-1$ & $+1$ & $0$ & $0$ \\
\hline
\end{tabular}
\end{center}
\begin{center}
\begin{tabular}{c}
\label{tab2}
Table (2)
\end{tabular}
\end{center}

In what follows, $L$ and $R$ refer to the projections of the Dirac
spinors into left and right handed parts, as in the standard model.
For a four component Dirac spinor $q$, we define $\overline{q} =
q^\dagger \gamma^0$ and
\begin{eqnarray}
\label{prime-op}
\begin{array}{lcl}
q_L& = & \frac{1}{2}(1+\gamma_5) q\\[0.05 in]
{q'}_L       & = & \gamma^2 (q_R)^\ast
\end{array}
& \hspace{0.05 in} &
\begin{array}{rcl}
q_R          & = & \frac{1}{2}(1-\gamma_5) q\\[0.05 in]
{q'}_R       & = & \gamma^2 (q_L)^\ast
\end{array}
\end{eqnarray}

The prime $(')$ operation changes the handedness of an object.
The reader is referred to the Appendix,  for definitions of the
$\gamma$'s.

Given a bilinear in (anti-commuting) Dirac spinors $q,r$ with
$c$-number coefficients $c_{\alpha\beta}$, we define the operation of
anti-symmetrized hermitianization  ($A.H.)$ as follows:
\begin{eqnarray}
\label{AHeqn}
\begin{array}{lll}
A.H. \; ( c_{\alpha\beta} q_\alpha r_\beta )  = 
\frac{1}{4}
\left[
c_{\alpha\beta} q_\alpha r_\beta +
\right.&&\\
\left. c_{\alpha\beta}^\ast (r_\beta)^\ast (q_\alpha)^\ast - c_{\alpha\beta} r_\beta q_\alpha -
c_{\alpha\beta}^\ast (q_\alpha)^\ast (r_\beta)^\ast
\right]&&
\end{array}
\end{eqnarray}

The interactions of the charged vector bosons with the leptons in the
family are given by the following terms in the Lagrangian (\ref{Lag}):

\begin{eqnarray}
\label{weakterms}
\begin{array}{lll}
 A.H.\frac{g}{\sqrt{2}}\left[ 
(
\begin{array}{cc}
\overline{{\nu}^{e_1}}_L & \overline{e}_{1L}
\end{array}
) \gamma^\mu \{
\left\lgroup
\begin{array}{cc}
0 & W_\mu^{1-}\\
W_\mu^{1+} & 0\\
\end{array}
\right\rgroup \right. && \\[0.15 in]
\left\lgroup
\begin{array}{c}
\nuleo\\
\eol
\end{array}
\right\rgroup
+ i W_\mu^{2-}
\left\lgroup
\begin{array}{c}
{e'}_{2L}\\
-{{\nu}'}^{e_2}_L
\end{array}
\right\rgroup
\}  &&\\[0.15in]
\left.
\phantom{[[[[}
+
(
\begin{array} {cc}
\overline{\nu^{e_2}}_R & \overline{e}_{2R}
\end{array}
) \gamma^\mu \{
\left\lgroup
\begin{array}{cc}
0 & W_\mu^{1-}\\
W_\mu^{1+} & 0\\
\end{array}
\right\rgroup \right. &&\\[0.15 in]
\left. \left\lgroup
\begin{array}{c}
\nuret\\
\etr
\end{array}
\right\rgroup
+ i W_\mu^{2-}
\left\lgroup
\begin{array}{c}
{e'}_{1R}\\
-{{\nu}'}^{e_1}_R
\end{array}
\right\rgroup
\}
\right]&&
\end{array}
\end{eqnarray}

The interactions of the neutral vector bosons with the leptons in the
family are given by the terms:

\begin{eqnarray*}
\frac{1}{2} \sqrt{f^2+g^2}[\tilde{Q}_{a}(q_h)+ cos  2\theta_w Q_{a}(q_h)] \cdot \overline{q}_h \gamma^\mu q_h {\cal Z}_\mu^a &&\\[0.1 in]
\mbox{where   } \; sin\; \theta_w = \frac{f}{\sqrt{f^2+g^2}}, \; \;  h = L, R\;\;\;\;\;\;\; &&
\end{eqnarray*}
and $Q_{a}(q_h)$ and $\tilde{Q}_{a}(q_h)$ are $c$-numbers tabulated in Table (4)),  page (\pageref{tab4}) for $q$ ranging 
over leptons and quarks.

The quark anti-quark interactions with massless gauge bosons (gluons \cite{Fritzsch, GrossW1, Politzer, Muta} ) and the 
associated gauge group is beyond the scope of the present communication which is restricted to the electroweak interactions
and phenomenological aspects of nuclear forces only.
\\[0.06 in]
\\
However it may be mentioned that the quark antiquark interactions are such that they give rise to protons $p,p'$ and
neutrons $n, n'$, all with baryonic number $B=1$.  When viewed in light
of the first or second kind (${\cal A}_\mu^1$, ${\cal A}_\mu^2$ resp.)
these composite particles have charges and designations as tabulated
in Table (3), page (\pageref{tab3}). This concept is illustrated in Fig.(1),
page (\pageref{arrow-table}) and Fig. (7), page (\pageref{tritium-fig}). 
\\
\\[0.05 in]

\begin{center}
\label{tab3}
\begin{tabular}{|c||c|c|c|c|}
\hline
composite & name given when & $Q_1$ & $Q_2$ & name given when \\
particle  & viewed with ${\cal A}_\mu^1$ & charge & charge & viewed with ${\cal A}_\mu^2$\\
& [ nomenclature &&&\\
&   used here]&&&\\
\hline
\hline
$u_2 d_2 u_1^\ast$      & $n$  & $0$ & $1$ & $p'$ \\
$u_2 d_2 d_1^\ast$      & $p$  & $1$ & $1$ & $p$  \\
$u_1^\ast d_1^\ast u_2$ & $p'$ & $1$ & $0$ & $n$  \\
$u_1^\ast d_1^\ast d_2$ & $n'$ & $0$ & $0$ & $n'$ \\
\hline
\end{tabular}\\
\end{center}
\begin{center}
\begin{tabular}{c}
\mbox{Table (3)}
\end{tabular}
\end{center}

Note that the designation of a composite particle is $proton$ when its
associated charge $Q_1= 1$, and $neutron$ when its associated charge $Q_1= 0$.
Conformity with the usual notation, in which the experimentally observed stable proton $p$ and unstable
neutron $n$ appear unprimed can not be maintained for all  scenarios discussed in the following. 
This is because, in this model there is another proton $p'$ and
another neutron $n'$, whose designations are primed [ not to be confused with the prime operation on Dirac spinors
defined by equation (\ref{prime-op}) ].

Three possible scenarios $ (Y, X_1, X_2) $ arise naturally in the model.

Scenario: $Y$: In this scenario it is assumed that $ m(u_1),\; m(d_1) << m(u_2),\; m(d_2)$ which implies that $m(p),\;m(n)>> 
m(p'),\; m(n')$. Comparison of the observed value for R (\ref{rvalue}) with the 
value calculated  in this scenario suggests that the quarks have ''color" and that the number of quark colors is 2. 
However this scenario violates some aspects of left-right symmetry in the model at a fundamental level which 
is against the ''{\em raison de etre}"  for the model. We briefly mention that models demonstrating confinement of two colors 
were constructed  some time ago \cite{Khan5, Khan6}. This is in striking contrast with the situation for three colors
which have never been shown to exhibit confinement. However we shall not present an investigation of this scenario in the 
present communication which deals with the $X$ scenarios described below.

Scenario $X_1$: In this scenario $m(n') = m(p)\; \mbox{and} \; m(p') =m(n)$. It arises from imposing a
symmetry constraint $m(u_1) = m(u_2)\;\mbox{and} \;m(d_1)=m(d_2)$ together with a constraint
on the values of $g,\; f$, and the independent mass ratios among the masses [$m(u_1),\; m(d_1),\;$
$ m(W),\; m(e),\;m(Z)$].

Scenario $X_2$: In this scenario $m(n) = m(n')\; \mbox{and} \; m(p) =m(p')$. 
It arises from imposing a
symmetry constraint $m(u_1) = m(d_2)\; \mbox{and}\; m(u_2)= m(d_1)$ together with another constraint
on the values of $g,\; f,$ and the independent mass ratios among the masses [$\;m(u_1),\; m(u_2),\;$
$ m(W),\; m(e),\;m(Z)\;$]. This scenario is in apparent disagreement with observations: see section (\ref{nuclei}) on
page (\pageref{scenx}) for a discussion of the $X$ scenarios.
\\
\\
The model satisfies the requirements for a renormalizable quantum field
theory \cite{Schweber, Bollini, Ashmore, Zuber}.  In particular, divergences due to anomalous Feynman diagrams 
(Adler-Bell-Jackiw anomalies) \cite{Adler, Bell, Bouchiat, GG, GrossJ} cancel.  These divergences cancel separately 
for the leptons and quarks, quite unlike the situation in the standard model. Therefore imposing the requirement
that these divergences cancel among the leptons and quarks in a single family do not restrict
the number of colors for quarks and leptons to any specific values. In the standard model, however, these
values are 3 and 1 respectively. Following the usual practice, we have to check that the experimentally measured ratio,
\begin{eqnarray}
\label{rvalue}
R& = &\frac{e_1^+ + e_1^- \longrightarrow \mbox{hadrons}}{e_1^+ + e_1^- \longrightarrow \mu^+_1 + \mu^-_1}
\end{eqnarray}

at center of mass energy $T$,
be in reasonable agreement with the value of $\sum_q Q_1^2(q)$, where $q$ ranges
over all leptons and quarks of mass $< T/2c^2$.  This
requirement constrains the
model (in scenarios $X_a$) to have one color for each of the quarks and leptons.
With this value for the number of quark colors, we find that when
$T$ lies between the threshold energies for $\tau$ lepton production
and bottom quark ($b$) production, the value of $\sum_q Q_1^2(q)$ is
$\frac{47}{9}$ (if we assume that production of only two charged neutrinos is effective in this energy region).
This value is in better agreement with the SLAC  data \cite{Huang1}, than the value found in the standard
model ($\frac{13}{3}$) - see \cite{Huang2}.

Finally a very interesting feature of the model is that it
predicts a charged neutrino, which is coupled to both kinds of
photons. This has experimental and theoretical consequences which are discussed in the section
(\ref{neutrino}).

\begin{figure}[htbp]
\centering{\mbox{\psfig{figure=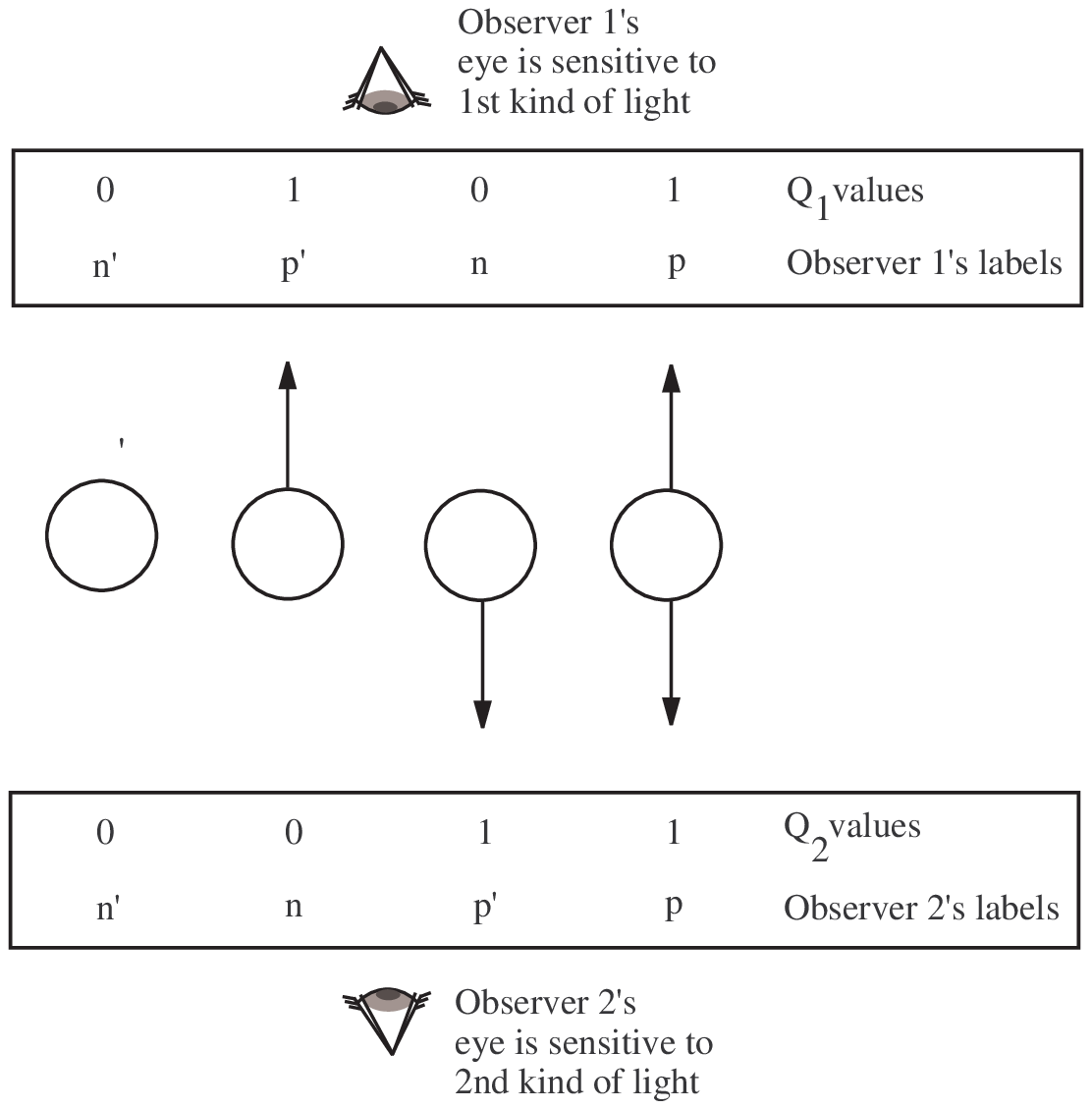}}}
\label{arrow-table}
\caption{Differing Perspectives
 of Observers}
\end{figure}

\section{Deductive Construction of the Model}
\label{section-3}
We present the construction in three stages, outlined here:
\begin{itemize}
\item
In stage (1) we construct the invariants under $R_1$ and $R_2$, but
ignore the groups  $U_1$ and $U_2$.
\item
Then, in stage (2) we introduce the two gauge groups of phase
transformations $U_1$ and $U_2$, and their associated gauge fields
$B_\mu^1$ and $B_\mu^2$.  Additional terms in the invariants
contructed in stage (1) are then deduced, so as to satisfy the
following three requirements:
\begin{enumerate}
\item
Each of the groups $U_1$ and $U_2$ commute with both the groups $R_1$
and $R_2$.
\item
Parity is conserved in interactions of each of the photons, i.e.
that 
$Q_a(q_{L}) = Q_a(q_{R})$,  where $a=1,2$ and $q$ varies over quarks and leptons.
\\
\item
All divergences in anomalous Feynman diagrams mutually cancel.
\end{enumerate}
\item

\item
Finally, in stage (3), section (\ref{stage-3}) we  review stages (1) and (2) in order to define those
aspects of the model pertaining to these stages which could not be specified until the 
developments in stage (2) were presented. At this final stage the compelling necessity of
pairs of families of quarks and leptons is established.

The discussion of the ''parity" transformation in section (\ref{parity}) on page (\pageref{parity})
corroborates the main results of the sections (\ref{stage-2}) and (\ref{stage-3}). 
\end{itemize}

\subsection{Stage (1)}
\label{stage-1}

We begin by considering the possible choices for the construction of invariants under $R_1$ and $R_2 $ 
in the model.  In what follows, a quark or lepton spinor projection with a definite handedness $
h \;=(L$ or $R)$ will be called an {\em object}.

\subsubsection{Construction of Invariants}

We first mention that for each choice of a gauge field $A_\mu^i$, from among the gauge
fields $(W_\mu^{a,i})$ introduced earlier, the transformations of $A_\mu^i$
occur  under the action of only one gauge group $R$, from among the gauge groups ($R_1$, $R_2$).
This group $R$ (with generators $R^i$) is that corresponding to the gauge field $A_\mu^i$, 
with coupling constant $g$.

[ Cautionary remark: $A^i_\mu$ and ${\cal{B}}^i_\mu$ are generic notations introduced 
in this section only and are not to be confused with the electromagnetic fields ${\cal{A}}^a_\mu$ and gauge fields
$B^a_\mu$ associated with the gauge groups $U_a$ , everywhere else. ]

The infinitesimal transformation $1+\theta^i R^i$ of $R$ acting on $A_\mu^i$ will be
assumed to always act as follows:
\begin{eqnarray}
A_\mu^i & \rightarrow & A_\mu^i + \epsilon^{ijk} \theta^j A_\mu^k + \frac{1}{g}\partial_\mu \theta^i
\end{eqnarray}
In contrast to this universal behavior of $A_\mu^i$, the transformations of Dirac spinors ( $q, r, s$
etc ) are determined by the various possible choices for  invariants given below. These invariants
involve {\em bilinears} of Dirac {\em spinors} and gauge fields. In each case of a possible invariant given in
the following, the corresponding actions of the infinitesimal transformations on the various spinors 
occurring in the invariant will also be given for a complete description of the invariant. 
A {\em basic invariant} for $R_1$, $R_2$ involves a pair of objects
$(q_h,r_h)$ which are either both quarks or both leptons (not necessarily 
of the same family and not necessarily both primed or both unprimed). The
members of the pair may have {\em different} handedness $h$.  This pair of objects
may be associated with any one of the gauge fields $(W_\mu^{a,i})$. For each choice of gauge field 
($A_\mu^j$) there is a unique possibility for constructing a basic invariant involving the pair of
leptons or quarks of definite handedness.

\begin{itemize}

\item[$1$] {Basic Invariant.}

The objects considered in this and the following subsection (3.1.2) have the same handedness [different handedness is possible also
- however, it is introduced only in the subsection on standard invariants (3.1.3) ]. Thus one or both the objects may be
primed objects with the same handedness (see equation (\ref{prime-op} ) for definition of primed object) i.e. the pair of 
objects is either
$ \left\lgroup
\begin{array}{c}
q_{h}\\
r_{h}
\end{array}
\right\rgroup
$
or
$
\left\lgroup
\begin{array}{c}
{q_{h}}\\
{r'_{h}}
\end{array}
\right\rgroup
$
or
$
\left\lgroup
\begin{array}{c}
{q'_{h}}\\
{r'_{h}}
\end{array}
\right\rgroup $.
In the first case (and similarly for the other cases), the invariant is 
\begin{eqnarray}
\label{basic-invariant1}
\begin{array}{rcl}
&& A.H.\left[ i
({\overline{q}}_h, {\overline{r}}_h)
(\gamma^\mu \partial_\mu - \right.\\
&&\left. \frac{ig}{2}{\sigma}^i(\xi_1, \xi_2)A_\mu^j)
\left\lgroup
\begin{array}{c}
q_h\\
r_h
\end{array}
\right\rgroup \right]
\end{array}
\end{eqnarray}
where 
\begin{eqnarray}
\begin{array}{lcl}
&&{\sigma}^i(\xi_1,\xi_2) =  [ \delta^i_2 + \xi_1(1  -  \delta^i_2)]{\tau}^{-1}(\xi_2) {\sigma}^i \tau(\xi_2)) \\[.15 in]
&& \tau(\xi_2)=
\left\lgroup
\begin{array}{cc}
\xi_2^* & 0\\
0 & \xi_2
\end{array}
\right\rgroup,\; \xi_1 =  \pm 1, \;|\xi_2| = 1 
\end{array}
\end{eqnarray}
[A.H. is defined by equation (\ref{AHeqn}).]

This is an invariant if we require that under the action of
the infinitesimal element $(1 + {\theta}^i R^i)$ of the group R, $q_{h}$ and $r_{h}$ 
transform as follows:
\begin{eqnarray}
 \left\lgroup
\begin{array}{c}
q_{h}\\
r_{h}
\end{array}
\right\rgroup
&
\rightarrow
&
[I + \frac{i}{2}{\theta}^i {\sigma}^i(\xi_1, \xi_2)]
\left\lgroup
\begin{array}{c}
q_{h}\\
r_{h}
\end{array}
\right\rgroup
\end{eqnarray}
 
We refer to the term (\ref{basic-invariant1}) as a {\em basic linkage } between $A_\mu^i$ and 
$(q_h, r_h)$.  This type of linkage will be depicted as
\addtocounter{figure}{1}
\begin{eqnarray*}
\begin{array}{c}
A_{\mu}[\xi_1,\xi_2] \left\{
\begin{array}{c}
q_h\\
\\
r_{h}
\end{array}
\right.\\
\\
\mbox{ Fig. (\arabic{figure}): Basic Linkage}
\end{array}
\end{eqnarray*}

when displaying the invariants in a pictorial representation of the Lagrangian.
\\
\end{itemize}
\begin{itemize}
\item[$2$] {Alternative form of Basic Invariant}

Another form of the basic invariant involving one gauge field and two objects of the same handedness ( one
or both of which may be primed) is now given. This form is more explicit and convenient to use.

\begin{eqnarray}
\label{basic-invariant2}
\begin{array}{l}
A.H. \left[ i \overline{q}_h \gamma^\mu
\left\{ \partial_\mu q_h + \frac{ig}{2}\eta_1 A_\mu^3 q_h \right.\right.\\[0.1 in]
\phantom{A.H.[}\left.-\frac{ig}{2}\eta_2( A_\mu^2 + i\eta_1 A_\mu^1) r_h \right\} \\[0.1in]
\phantom{A.H.[}
\left.
+ i \overline{r}_h \gamma^\mu
\left\{ \partial_\mu r_h - \frac{ig}{2}\eta_1 A_\mu^3 r_h \right.\right.\\[0.1 in]
\phantom{A.H.[}\left.\left.-\frac{ig}{2}\eta_2^*( A_\mu^2 - i\eta_1 A_\mu^1) q_h \right\}
\right]
\end{array}
\end{eqnarray}

\begin{eqnarray*}
&& \mbox{where} \; \eta_1=\xi_1,\;   \eta_2= -i ({\xi}_2)^2\\
\\
&& \mbox{so that}\; \eta_1 = \pm 1,\;  |\eta_2| = 1
\end{eqnarray*}

and the associated transformations of $q_h$ and ${r}_h$ under an
infinitesimal element $(1 + \theta^i R^i)$ of R are given by
\begin{eqnarray}
\label{variation1}
\begin{array}{rcl}
&& q_h \rightarrow q_h - \frac{i}{2}\eta_1 \theta^3  q_h\\
&& \phantom{q_h \rightarrow} +\frac{i}{2} \eta_2(\theta^2 + i \eta_1
\theta^1){r}_h\\
&& r_h \rightarrow {r}_h + \frac{i}{2}\eta_1 \theta^3 {r}_h \\
&&\phantom{r_h \rightarrow}+\frac{i}{2} {\eta_2}^*(\theta^2 -i \eta_1
\theta^1)q_h
\end{array}
\end{eqnarray}

The invariant linkage (\ref{basic-invariant2}) between $A_\mu^i$ and $(q_h, r_h)$
will be depicted graphically as follows:
\\
\\
\addtocounter{figure}{1}
\begin{eqnarray*}
\begin{array}{c}
A_{\mu}[\xi_1,\xi_2] \left[
\begin{array}{c}
q_h\\
\\
r_{h}
\end{array}
\right.\\
\\
\\
\\
\mbox{ Fig. (\arabic{figure}): Alternative Form of Basic Linkage}
\end{array}&&
\end{eqnarray*}
\end{itemize}

Parenthetical remarks:

1) Notice that the equivalence of the two forms of the basic invariant ( \ref{basic-invariant1} and \ref{basic-invariant2})
introduced above is expressed by

\begin{eqnarray*}
\begin{array}{c}
A_{\mu}[\xi_1,\xi_2] \left\{
\begin{array}{c}
q_h\\
\\
r_{h}
\end{array}
\right.
\end{array}&=& 
\begin{array}{c}
A_{\mu}[\eta_1,\eta_2] \left[
\begin{array}{c}
q_h\\
\\
r_{h}
\end{array}
\right.\\
\end{array}
\end{eqnarray*}
\begin{eqnarray*}
\mbox{where}\; & \eta_1=\xi_1, & \eta_2= -i ({\xi}_2)^2
\end{eqnarray*}
2) Notice that variations (\ref{variation1}) imply

\begin{eqnarray}
\label{variation2}
\begin{array}{rcl}
&&{q'}_{\tilde{h}} \rightarrow {q'}_{\tilde{h}} + \frac{i}{2}\eta_1 \theta^3 {q'}_{\tilde{h}}\\
&&\phantom{{q'}_{\tilde{h}} \rightarrow} -\frac{i}{2} {\eta_2}^*(\theta^2 - i \eta_1
\theta^1)r'_{\tilde{h}}\\
&&{r'}_{\tilde{h}} \rightarrow {r'}_{\tilde{h}} - \frac{i}{2} \eta_1 \theta^3  {r'}_{\tilde{h}}\\
&&\phantom{{r'}_{\tilde{h}} \rightarrow} -\frac{i}{2} \eta_2(\theta^2 +i \eta_1
\theta^1){q'}_{\tilde{h}}
\end{array}
\end{eqnarray}

where $h = L\;\mbox{or}\; R \;\mbox{and}\; \tilde{L}=R , \; \tilde{R}=L $

Also, because of the equivalence of the variations (\ref{variation1}) and (\ref{variation2}),
the previously defined graphic notation allows us to write:
\begin{eqnarray*}
A_{\mu}[\eta_1,\eta_2] \left[
\begin{array}{c}
q_h\\
\\
r'_{h}
\end{array}
\right.
&=& 
A_{\mu}[-\eta_1,-{{\eta}_2}^*] \left[
\begin{array}{c}
q'_h\\
\\
r_{h}
\end{array}
\right.\\
\end{eqnarray*}
\begin{eqnarray*}
A_{\mu}[\eta_1,\eta_2] \left[
\begin{array}{c}
q'_h\\
\\
r'_{h}
\end{array}
\right.
&=& 
A_{\mu}[-{\eta}_1,-{{\eta}_2}^*] \left[
\begin{array}{c}
q_h\\
\\
r_{h}
\end{array}
\right.\\
\end{eqnarray*}

\subsubsection{Extended Invariants}
The invariant next in the order of length is now described.
An {\em extended invariant} for the commuting groups $R_1$, $R_2$ involves 
{\em two} pairs of objects $(q_h,r_h)$, and $(q_h,s_h)$ having one
member in common.  All objects involved in an extended invariant may have 
{\em different} handedness. In the following description of the invariant,however,
we shall take them to have the same handedness $h$ and they are either all quarks
or all leptons.  In each pair, one or both objects may be primed ($'$) or unprimed.

Each pair of objects may be associated with any one of the gauge fields
$(W_\mu^{a,j}, a=1, 2 )$ introduced earlier, so long as the two pairs
are associated with different gauge fields.  For each choice of
distinct gauge fields $A_\mu^i$ and ${\cal{B}}_\mu^i$ , associating $A_\mu^i$
with $(q_h,r_h)$ and ${\cal{B}}_\mu^i$ with $(q_h,s_h)$ we get the
following invariant.
\begin{eqnarray}
\label{extended-invariant}
\begin{array}{l}
A.H. \left[
i {\overline{q}}_h \gamma^\mu
\left\{
( \partial_\mu + \frac{ig}{2} \xi_1 A^3_\mu )
q_h \right. \right. \\[0.1in]
- \frac{ig}{2}\xi_2 ( A^2_\mu + i \xi_1 A^1_\mu ) {r_h}\\[0.1 in]
\left.
+  \frac{ig}{2}\eta_1 {\cal{B}}^3_\mu q_h   
-  \frac{ig}{2} \eta_2
({\cal{B}}^2_\mu + i \eta_1 {\cal{B}}^1_\mu) {s_h}\right\}\\[0.1 in]
+i \overline{r}_h \gamma^\mu
\left\{ (\partial_\mu -  \frac{ig}{2}\xi_1 A^3_\mu) {r_h}
 \right.\\[0.1in]
\left. -  \frac{ig}{2} {\xi}_2^*(A^2_\mu - i \xi_1 A^1_\mu) q_h
\right\}\\[0.1 in]
 + i \overline{s}_h \gamma^\mu \left\{
(\partial_\mu - \frac{ig}{2} \eta_1 {\cal{B}}_\mu^3) {s_h} \right. \\[0.1 in]
\left. \left. -  \frac{ig}{2} {\eta}_2^*
({\cal{B}}_\mu^2 - i \eta_1     {\cal{B}}_\mu^1) q_h\right\}
\right]
\end{array}
\end{eqnarray}
\begin{eqnarray*}
&\mbox{where }\; \xi_1,\; \xi_2,\; \eta_1,\; \eta_2 \;\mbox{are  independent,}&\\
&\xi_1 = \pm 1,\;|\xi_2| = 1,\;\eta_1 = \pm 1,\;|\eta_2| = 1.&
\end{eqnarray*}
The associated transformations of $q_h$, ${r}_h$ and ${s}_h$ under an
infinitesimal element $(1 + \theta^i_1 T^i + \theta^i_2 S^i)$ of $R_1$ and $R_2$
( with generators $T^i$ and $S^i$ ) are given by
\begin{eqnarray}
\label{variation3}
\begin{array}{lcl}
q_h \rightarrow q_h - \frac{i}{2}\xi_1 \theta^3_1  q_h +\frac{i}{2} \xi_2(\theta^2_1 + i \xi_1
\theta^1_1){r}_h &&\\[0.1 in]
\;\;- \frac{i}{2} \eta_1 \theta^3_2  q_h +\frac{i}{2} \eta_2(\theta^2_2 + i \eta_1
\theta^1_2){s}_h&&\\[0.1 in]
{r}_h  \rightarrow {r}_h + \frac{i}{2}\xi_1 \theta^3_1 {r}_h +\frac{i}{2} {\xi_2}^*(\theta^2_1 -i \xi_1
\theta^1_1)q_h&&\\[0.1 in]
{s}_h  \rightarrow {s}_h + \frac{i}{2} \eta_1 \theta^3_2  {s}_h +\frac{i}{2} {\eta_2}^*(\theta^2_2 -i \eta_1
\theta^1_2)q_h&&
\end{array}
\end{eqnarray}
We refer to the term (\ref{extended-invariant}) as an {\em extended linkage} between $A_\mu^i$, $(q_h, {r_h})$ and ${\cal{B}}_\mu^i$,
$(q_h, {s_h})$.

When displaying the invariants in the Lagrangian, this type of linkage will be depicted in the manner shown in Figure(4).

\addtocounter{figure}{1}
\begin{eqnarray*}
&\begin{array}{c}
{\cal{B}}_{\mu}[\eta_1, \eta_2] \left[
{\begin{array}{c}
A_{\mu}[\xi_1, \xi_2] \left[
\begin{array}{c}
\;\;q_h
\\
\\
\;\;r_h
\end{array}
\right.
\\
\\
\\
\end{array}
\negthickspace\negthickspace\negthickspace\negthickspace\negthickspace\negthinspace
}{\begin{array}{c}
\\
\\
\\
\\
s_h
\end{array}}
\right.
\end{array}
&\\
&&\\
&\mbox{Figure (\arabic{figure}): Extended Linkage}&\\
\end{eqnarray*}
[Cautionary remark: Notice that the transformations of the objects in extended invariants are different from  
those of the objects in basic invariants. Therefore one must not combine basic and extended invariants
into the same Lagrangian if the two kinds of invariants have objects in common.]

\subsubsection{Standard Invariants}

The Lagrangian of elementary particle interactions according to the present model is composed exclusively of
standard invariants defined in the following.
A {\em standard invariant} for $R_1$, $R_2$ involves a {\em doublet of pairs}: i.e. one pair of  objects
$(q_h,r_{h'})$ (which may have different handedness $ h, {h'}$) and another pair of objects with handedness opposite to 
that of the first pair, viz. $(s_{\tilde{h}},t_{\tilde{{h'}}})$, where $q,r,s,t$ are either all leptons or all quarks.

Each doublet of ordered pairs above may be associated with any ordered pair $(A_\mu^i, {\cal{B}}_\mu^i)$ of distinct 
gauge fields from the $(W_\mu^{a,i}$). Each such association gives rise to an invariant. One of the two possible
invariants thus arising is given below.

\begin{eqnarray}
\label{standard-invariant1}
\begin{array}{l}
A.H. \left[
i (
\begin{array}{cc}
{\overline{q}}_h& {\overline{r}}_{h'}
\end{array}
)
\gamma^\mu
\{
\partial_\mu
\left\lgroup
\begin{array}{l}
q_h\\
r_{h'}
\end{array}
\right\rgroup
\right.\\[0.15 in]
-
\frac{ig}{2} \sigma^j (\xi_1,\xi_2)A_\mu^j
\left\lgroup
\begin{array}{c}
q_h\\
r_{h'}
\end{array}
\right\rgroup 
\left.
- \frac{ig}{2}\eta_1 {\cal{B}}_\mu^3
\left\lgroup
\begin{array}{c}
q_h\\
r_{h'}
\end{array}
\right\rgroup \right.\\
 - \frac{ig}{2} \eta_2 ( {\cal{B}}_\mu^2 + i\eta_1 {\cal{B}}_\mu^1 )
\left\lgroup
\begin{array}{c}
t'_{h'}\\
-s'_h
\end{array}
\right\rgroup \}                            \\[0.15in]
\phantom{
A.H. \left[\right.}
+ i (
\begin{array}{cc}
\overline{s}_{\tilde{h}} & \overline{t}_{\tilde{h'}}
\end{array}
)
\gamma^\mu
\{
\partial_\mu
\left\lgroup
\begin{array}{c}
s_{\tilde{h}}\\
t_{\tilde{h'}}
\end{array}
\right\rgroup
        \\[0.15in]
\left.
\left.
- \frac{ig}{2} \sigma^j(\xi_1,\xi_2) A_\mu^j
\left\lgroup
\begin{array}{c}
s_{\tilde{h}}\\
t_{\tilde{h'}}
\end{array}
\right\rgroup 
- \frac{ig}{2}\eta_1 {\cal{B}}^3_\mu 
\left\lgroup
\begin{array}{c}
s_{\tilde{h}}\\
t_{\tilde{h'}}
\end{array}
\right\rgroup \right. \right.\\
\left.\left. - \frac{ig}{2} \eta_2
( {\cal{B}}_\mu^2 + i\eta_1 {\cal{B}}_\mu^1 )
\left\lgroup
\begin{array}{c}
{r'}_{\tilde{h'}}\\
-{q'}_{\tilde{h}}
\end{array}
\right\rgroup
\}
\right]\right.
\end{array}
\end{eqnarray}

This linkage is represented diagrammatically as follows:

\addtocounter{figure}{1}
\begin{eqnarray*}
{\cal{B}}_{\mu}[\eta_1,\eta_2] \left[
\begin{array}{c}
A_{\mu}[\xi_1,\xi_2] \left\{
\begin{array}{c}
q_h\\
\\
r_{h'}
\end{array}
\right.\\
\\
A_{\mu}[\xi_1,\xi_2]\left\{
\begin{array}{c}
s_{\tilde{h}}\\
\\
t_{\tilde{h'}}
\end{array}
\right.
\end{array}
\right.
&&\\
&&\\
&&\\
\mbox{Figure \arabic{figure}: Standard Linkage}
\end{eqnarray*}

The associated transformations of $q_h$, $r_{h'}$, $s_{\tilde{h}}$ and $t_{\tilde{h'}}$ under an
infinitesimal element $(1 + \theta^i_1 T^i + \theta^i_2 S^i)$ of $R_1$ and $R_2$
are given by

\begin{eqnarray}
\begin{array}{lcl}
\label{variation4}
\left\lgroup
\begin{array}{c}
q_h\\
r_{h'}
\end{array}
\right\rgroup
\rightarrow 
\left\lgroup
\begin{array}{c}
q_h\\
r_{h'}
\end{array}
\right\rgroup
+\frac{i}{2}\theta^i_1 \sigma^i(\xi_1,\xi_2)
\left\lgroup
\begin{array}{c}
q_h\\
r_{h'}
\end{array}
\right\rgroup&&\\[0.15 in]
 + \frac{i}{2}\eta_1 \theta^3_2
\left\lgroup
\begin{array}{c}
q_h\\
r_{h'}
\end{array}
\right\rgroup
 + \frac{i}{2} \eta_2(\theta^2_2 + i \eta_1
\theta^1_2)
\left\lgroup
\begin{array}{c}
t'_{h'}\\
-s'_h
\end{array}
\right\rgroup
&&\\[0.15 in]
&&\\[0.15 in]
\left\lgroup
\begin{array}{c}
s_{\tilde{h}}\\
t_{\tilde{h'}}
\end{array}
\right\rgroup
\rightarrow 
\left\lgroup
\begin{array}{c}
s_{\tilde{h}}\\
t_{\tilde{h'}}
\end{array}
\right\rgroup
+\frac{i}{2}\theta^i_1 \sigma^i(\xi_1,\xi_2)
\left\lgroup
\begin{array}{c}
s_{\tilde{h}}\\
t_{\tilde{h'}}
\end{array}
\right\rgroup &&\\[0.15 in]
 + \frac{i}{2}\eta_1 \theta^3_2
\left\lgroup
\begin{array}{c}
s_{\tilde{h}}\\
t_{\tilde{h'}}
\end{array}
\right\rgroup
 + \frac{i}{2} \eta_2(\theta^2_2 + i \eta_1
\theta^1_2)
\left\lgroup
\begin{array}{c}
r'_{\tilde{h'}}\\
-q'_{\tilde{h}}
\end{array}
\right\rgroup
&&
\end{array}&&
\end{eqnarray}

We refer to the term (\ref{standard-invariant1}) as a {\em Standard Linkage} (SL) 
between $A_\mu^j, {\cal{B}}_\mu^j$, $(q_h,r_{h'})$, and $(s_{\tilde{h}},t_{\tilde{h'}})$.

Notice that when $h \neq h'$ the invariant (\ref{standard-invariant1}) reduces to become

\begin{eqnarray}
\label{standard-invariant4}
\begin{array}{l}
A.H. \left[
i (
\begin{array}{cc}
{\overline{q}}_h& {\overline{r}}_{h'}
\end{array}
)
\gamma^\mu
\{
\partial_\mu
\left\lgroup
\begin{array}{c}
q_h\\
r_{h'}
\end{array}
\right\rgroup \right.\\[0.15 in]
\left.
-
\frac{ig}{2} \sigma^3 (\xi_1,\xi_2)A_\mu^3
\left\lgroup
\begin{array}{c}
q_h\\
r_{h'}
\end{array}
\right\rgroup 
- \frac{ig}{2}\eta_1 {\cal{B}}_\mu^3
\left\lgroup
\begin{array}{c}
q_h\\
r_{h'}
\end{array}
\right\rgroup
\}\right. \\[0.15in]
+ i (
\begin{array}{cc}
{\overline{s}}_{\tilde{h}} & {\overline{t}}_{\tilde{h'}}
\end{array}
)
\gamma^\mu
\{
\partial_\mu
\left\lgroup
\begin{array}{c}
s_{\tilde{h}}\\
t_{\tilde{h'}}
\end{array}
\right\rgroup \\[0.15 in]
\left.\left. - \frac{ig}{2} \sigma^3(\xi_1,\xi_2) A_\mu^3
\left\lgroup
\begin{array}{c}
s_{\tilde{h}}\\
t_{\tilde{h'}}
\end{array}
\right\rgroup  
- \frac{ig}{2}\eta_1 {\cal{B}}^3_\mu 
\left\lgroup
\begin{array}{c}
s_{\tilde{h}}\\
t_{\tilde{h'}}
\end{array}
\right\rgroup\}
\right]\right.
\end{array}
\end{eqnarray}

See equation (\ref{appendeq}) in the Appendix, page (\pageref{appendix}).

Given a linkage $L$, which is of any one of the previously described types, we define
its {\em domain} $\mathbf{D}(L)$ to be the set of quarks or leptons
associated with $L$.

We can now describe the first two terms in the Lagrangian  which are invariant under $R_1$ and $R_2$. These
four terms arise from  Standard Linkages between
\begin{eqnarray*}
W_\mu^{1j}, W_\mu^{2j}, \;({\nu}^{e_1}_L, e_{1L})&\; \mbox{ and }\; &({\nu}^{e_2}_R, e_{2R});\\ 
W_\mu^{1j}, W_\mu^{2j}, \;(u_{1L}, d_{1L})& \;\mbox{ and }\; & (u_{2R}, d_{2R}).
\end{eqnarray*}
All these linkages have the values for the parameters $\xi_1, \; \xi_2 \; \eta_1 \;
\eta_2$ all set equal to +1 as depicted in the diagrammatic representation of the Lagrangian (\ref{Lag}).

We now mention that at this first stage of  construction of the model the
transformation properties of the multiplets $( {\nu}^{e_1}_R, e_{1R}, {\nu}^{e_2}_L, e_{2L})$                     
and $(u_{1R}, d_{1R}, u_{2L}, d_{2L})$ under the actions of 
the generators of $R_1$ and $R_2$ have not been specified. This will be done in section (\ref{stage-3}) after we
become familiar with the contents of sections (\ref{stage-2}).

\subsection{Stage (2)}
\label{stage-2}

In this stage, we introduce two gauge groups of phase
transformations $U_1$ and $U_2$, together with their associated
gauge fields $B_\mu^1$ and $B_\mu^2$.  The linkage terms
described in the previous section (\ref{stage-1}) are now augmented so as to make them
invariant under $U_1$ and $U_2$, while satisfying the three
conditions formulated in section (\ref{section-3}), paragraph
 (3), page (\pageref{section-3}).

{\em First,} we require that each of the groups $U_1$ and $U_2$ commute
with both the groups $R_1$ and $R_2$.  This requirement can be
satisfied by adding the following term to each previously described linkage $L$:
\begin{eqnarray}
\label{augmentedby}
f
\negthickspace\negthickspace\negthickspace\negthickspace\negthickspace
\sum_{
\begin{array}{c}
{\scriptstyle q_h \text{ in } \mathbf{D}(L)}\\
{\scriptstyle a=1,2}
\end{array}}
\negthickspace\negthickspace\negthickspace\negthickspace\negthickspace
c_{a}(q_h) \cdot \overline{q}_h \gamma^\mu q_h B^a_\mu
\end{eqnarray}
(Here the $c_{a}(q_h)$ are $c$-numbers.)  In particular, if $L$ is a basic linkage or an extended linkage ,
then the $c_{a}(q_h)$ do not depend on  $q_h$.  In contrast, for a standard linkage ,
between say $(q_L,r_L)$ and $(s_R,t_R)$, the $c_{a}(q_h)$ have a dependence on $q_h$
expressed by
\begin{eqnarray}
\begin{array}{rcl}
\label{standard-inv-condition}
c_a({q_L}) = c_a({r_L}) = -c_a({s_R}) = -c_a({r_R})&&
\end{array}
\end{eqnarray} 
We refer to the invariants obtained by adding term (\ref{augmentedby}) to
each of the terms (\ref{basic-invariant1}), (\ref{basic-invariant2}), (\ref{extended-invariant}),
(\ref{standard-invariant1})
as {\em augmented linkages} of various types. The augmented standard linkages are represented diagrammatically
as follows:

\addtocounter{figure}{1}
\begin{eqnarray*}
{\cal{B}}_{\mu}[\eta_1,\eta_2] \left[
\begin{array}{c}
A_{\mu}[\xi_1,\xi_2] \left\{
\begin{array}{c}
\begin{array}{ccc}
q_h & c_1(q_h) & c_2(q_h)
\end{array}\\
\\
\begin{array}{ccc}
r_{h'}\ & c_1(r_{h'})& c_2(r_{h'})
\end{array}
\end{array}
\right.\\
\\
A_{\mu}[\xi_1,\xi_2]\left\{
\begin{array}{c}
\begin{array}{ccc}
s_{\tilde{h}}&  c_1(s_{\tilde{h}}) & c_2(s_{\tilde{h}})
\end{array}\\
\\
\begin{array}{ccc}
t_{\tilde{h'}} & c_1(t_{\tilde{h'}})& c_2(t_{\tilde{h'}})
\end{array}
\end{array}
\right.
\end{array}
\right.
&&\\
&&\\
&&\\
\mbox{Fig.(\arabic{figure}):Augmented Standard Linkage}&&\\
\end{eqnarray*}

{\em Second,} we must ensure that parity is conserved in interactions of each of the photons.
We know from experimental evidence that the second photon ${\cal A}_\mu^2$ does not interact with
the electron $e_1$---so we expect, by symmetry, that the photon ${\cal A}_\mu^1$
does not interact with the electron $e_2$.  This suggests that we define the
photons in the model to be orthogonal to

$
gW_\mu^{a,3} - f B_\mu^a \; \; \; \;  (a=1,2)
$

Hence,

\begin{eqnarray}
\label{A-mu-define-2}
\begin{array}{rcl}
{\cal {A}}_\mu^a & = & sin \;{\theta}_w \; W_\mu^{a,3} + cos \;{\theta}_w\; B_\mu^a 
\end{array}
\end{eqnarray} 

where 

$
sin\;{\theta}_w = \frac{f}{\sqrt{f^2+g^2}}\\
cos\;{\theta}_w = \frac{g}{\sqrt{f^2+g^2}}
$

\begin{eqnarray*}
\begin{array}{|c||c|c|c|c|c|c|c|c|}
\hline
&&&&&&&&\\
&&&&&Q_1 =& Q_2 =&{\tilde{Q}}_1=&{\tilde{Q}}_2=\\
q_h       &  c_1(q_h) &  c_2(q_h) &  c_3(q_h) &  c_4(q_h) &  c_5(q_h) &  c_6(q_h) &  c_7(q_h) &  c_8(h) \\
\hline
\hline
{\nu}^{e_1}_{L}  &  -1/2 & -1/2 & 1/2  &  1/2 &    0  &    0 &    1 &    1 \\
e_{1L}    &  -1/2 & -1/2 & -1/2 &  1/2 &   -1  &    0 &    0 &    1 \\
{\nu}^{e_1}_{R}  &  -1/2 & -1/2 & 1/2  &  1/2 &    0  &    0 &    1 &    1 \\
e_{1R}    &  -1/2 & -1/2 & -1/2 &  1/2 &   -1  &    0 &    0 &    1 \\
{\nu}^{e_2}_{R}  &   1/2 &  1/2 & 1/2  &  1/2 &    1  &    1 &    0 &    0 \\
e_{2R}    &   1/2 &  1/2 & -1/2 &  1/2 &    0  &    1 &   -1 &    0 \\
{\nu}^{e_2}_{L}  &   1/2 &  1/2 & 1/2  &  1/2 &    1  &    1 &    0 &    0 \\
e_{2L}    &   1/2 &  1/2 & -1/2 &  1/2 &    0  &    1 &   -1 &    0 \\
u_{1L}    &  -1/6 & -1/6 & 1/2  &  1/2 &  1/3  &  1/3 &  2/3 &  2/3 \\
d_{1L}    &  -1/6 & -1/6 & -1/2 &  1/2 &  -2/3 &  1/3 & -1/3 &  2/3 \\
u_{1R}    &  -1/6 & -1/6 & 1/2  &  1/2 &  1/3  &  1/3 &  2/3 &  2/3 \\
d_{1R}    &  -1/6 & -1/6 & -1/2 &  1/2 &  -2/3 &  1/3 & -1/3 &  2/3 \\
u_{2R}    &   1/6 &  1/6 & 1/2  &  1/2 &  2/3  &  2/3 &  1/3 &  1/3 \\
d_{2R}    &   1/6 &  1/6 & -1/2 &  1/2 &  -1/3 &  2/3 & -2/3 &  1/3 \\
u_{2L}    &   1/6 &  1/6 & 1/2  &  1/2 &  2/3  &  2/3 &  1/3 &  1/3 \\
d_{2L}    &   1/6 &  1/6 & -1/2 &  1/2 &  -1/3 &  2/3 & -2/3 &  1/3 \\
\hline
\end{array}
\end{eqnarray*}
\begin{center}
\begin{tabular}{c}
\label{tab4}
Table (4)
\end{tabular}
\end{center}

$\theta_w$ is the analogue of the angle $\theta_W$ in the standard model \cite{Weinberg1}.
We further define 

${\cal {Z}}_\mu^a = cos\; {\theta}_w\; W_\mu^{a,3} - sin \;{\theta}_w \;B_\mu^a $
\\
\\
 Re-expressing the
augmented linkages in terms of the just defined ${\cal A}_\mu^a$ and ${\cal {Z}}_\mu^a$, it is easy to verify that
 ${\cal A}_\mu^a$ only occurs in terms of the form $\frac{1}{2}\sqrt{f^2+g^2}\;sin\; 2\theta_w \; Q_{a}(q_h)\cdot
\overline{q}_h \gamma^\mu q_h {\cal A}_\mu^a$,
and ${\cal Z}_\mu^a$ occurs exclusively in terms of the form
$\frac{1}{2}\sqrt{f^2+g^2}( \tilde{Q}_{a}(q_h)+ cos \; 2\theta_w \;Q_{a}(q_h)) \cdot \overline{q}_h 
\gamma^\mu q_h {\cal Z}_\mu^a$.
[Here $Q_{a}(q_h)$ and $\tilde{Q}_{a}(q_h)$ are $c$-numbers tabulated in Table (4) for a Standard Linkage between 
$W_\mu^{1j}, W_\mu^{2j}$, $({\nu}^{e_1}_{L(R)}, e_{1L(R)})$ and
$({\nu}^{e_2}_{R(L)}, e_{2R(L)}) $ respectively and a Standard Linkage between $W_\mu^{1j}, W_\mu^{2j}$, 
$(u_{1L(R)}, d_{1L(R)})$ and $(u_{2R(L)}, d_{2R(L)})$ respectively.]

Parity conservation, whose precise notion in the model is given in section (\ref{parity}) page 
(\pageref{parity}),  is implemented by setting
\begin{eqnarray}
\label{ql=qr-condition}
& Q_{a}(q_L)  = Q_{a}(q_R) &
\end{eqnarray}
where $a=1,2$ and $q$ varies over quarks and leptons.

{\em Finally,} we require that all divergences in anomalous Feynman diagrams \cite{Bouchiat, GG, GrossJ} mutually cancel.
This implies that all matrices $V_1, V_2, (A)$ associated with vector 
(axial vector) bilinears  occurring at
vertices
in triangular Feynman diagrams with external lines of gauge bosons in the model must satisfy: 

\begin{eqnarray*}
Tr \left[ (V_1 V_2 + V_2 V_1), A\right] & = & 0
\end{eqnarray*}
which in turn implies that for all $a, b, c = 1,2$
\begin{eqnarray}
\label{qabc-condition}
\begin{array}{rcl}
\sum_q Q_{a}(q_h) Q_{b}(q_h) \left( \tilde{Q}_c(q_L) - \tilde{Q}_c(q_R) \right) = 0&&
\end{array}
\end{eqnarray}

Parenthetical remark: [We shall see later that according to the precise notion of ''parity symmetry" in the model introduced 
in section (\ref{parity}), page (\pageref{parity}) we have 
$
{\tilde{Q}}_{a}(q_L) = {\tilde{Q}}_{a}(q_R).
$
Thus the conditions ( \ref{qabc-condition}) implied by the cancellation of infinities in anomalous Feynman diagrams
\cite{Bouchiat, GG, GrossJ} are guaranteed when ''parity symmetry" for the model is implemented.
However at this stage we need not invoke the complete notion of ''parity symmetry" to be introduced later.]

In order to be in agreement with the specifications in the standard model,
we take
\begin{eqnarray}
\label{q1-e,u,d}
\begin{array}{rcl}
Q_1(e_1) = -1, \;\; \; \; Q_1(u_1) - Q_1(d_1) = +1&&
\end{array}
\end{eqnarray}
The constraints (\ref{standard-inv-condition}), (\ref{A-mu-define-2}), (\ref{ql=qr-condition}), (\ref{qabc-condition}), 
and (\ref{q1-e,u,d}) along with the fundamental requirement that baryons of integral charge and composed of three quarks
be allowed in the model then {\em uniquely} determine the values of the the coefficients, $c_i(q_h)\; \;(i=1-4)$
of the diagonal terms ${\overline{q}}_h{\gamma}^{\mu}q_h V_{\mu}^i$ appearing in the Lagrangian. $V_{\mu}^i$ ranges 
over $ fB_{\mu}^1,\; fB_{\mu}^2, \;gW_{\mu}^{13},\; gW_{\mu}^{23}$ which are
denoted, respectively, by $V^i_{\mu}$  $(i=1-4)$. These values $c_i(q_h)$, deduced for members of the first family 
under consideration, are tabulated in Table (4), page (\pageref{tab4}). In Table (4), $c_5,\; c_6,\; c_7,\; c_8$ are
defined as follows:
\begin{eqnarray*}
c_5=c_3+c_1, \;\; c_6=c_4+c_2 &&\\
c_7=c_3-c_1,\; \; c_8=c_4-c_2 &&
\end{eqnarray*}

Notice that at  stage (1) we did not specify the transformations under the actions of 
$R_1, \;R_2$ for the multiplets $( {\nu}^{e_1}_R, e_{1R}, {\nu}^{e_2}_L, e_{2L})$ and $(u_{1R}, d_{1R}, u_{2L}, d_{2L})$.
However we have just determined the actions of $U_1$ and $U_2$ on these multiplets which are specified by the
values of $c_1$ and $c_2$ given in  Table (4). We shall take the constraint provided by the actions of $U_1$ and $U_2$ on the
multiplets into account in stage (3) to determine the actions of $R_1$ and $R_2$ on these multiplets. The latter set of
actions must commute with the former.

\subsection{Stage (3)}
\label{stage-3}

In this final stage completing the construction of the part of the model not involving Higgs scalars \cite{Higgs1, Higgs2, Higgs3,
Ander1, Brout}, we would determine the transformation properties
under $R_1,\; R_2,$ of the multiplets ($ {{\nu}^{e_1}}_{R}, {e}_{1R},$ , 
$ {{\nu}^{e_2}}_L, {e}_{2L}$), $( u_{1R}, d_{1R},
u_{2L}, d_{2L})$ and the associated invariant terms in the Lagrangian.

To do this, we refer to the following guidelines:

1) The actions of $U_a$  (with infinitesimal generators $1 + i {\varepsilon}_a U^a$) on the multiplets were found in 
section(\ref{stage-2}, \pageref{stage-2} to be given by

$
q_h \rightarrow q_h  - i c_a(q_h){\varepsilon}_a q_h
$

where $c_a (q_h)$ are tabulated in Table (4).

2) The actions just mentioned in (1) must commute with the actions of $R_1$ and $R_2$ to be found.

3) The actions of $R_1$ and $R_2$ are determined by the choice of invariants linking the multiplets. This choice must be
such that it reproduces the specific feature of the standard model involving $e_{1R}$ (the right-handed projection of
the usual electron): viz., $e_{1R}$ is not coupled to the charged
vector bosons $W_{\mu}^{a \pm}$ since there are no right-handed weak currents manifested in the phenomenology of weak 
interactions involving electrons \cite{Riazuddin}.

It can be shown that it is not possible to satisfy the three guidelines given above within the framework of the 
family of leptons and quarks introduced so far. However, it is possible to do so if we introduce 
a new family of leptons and quarks for which the set of values $c_j(q_h),\; [j=1,2]$  are
equal to those of the first family of leptons and quarks [given in Table (4)].
We designate the members of the new family by the addition of a ''breve" to the symbols for the
corresponding members of the first family of quarks and leptons:  viz.,    $({\breve{{\nu}}}^{{e}_1}, {\breve{e}}_{1},
 {\breve{\nu}}^{{e}_2},
{\breve{e}}_{2})$ and $({\breve{u}}_1, {\breve{d}}_1, {\breve{u}}_2, {\breve{d}}_2)$. The ''breve" family's members 
are arranged 
in their interactions with the gauge
bosons analogously to the already described partial organization of the first family, except that the roles of left-handed
and right-handed objects are interchanged (for the reason to emerge below).
Thus the two terms in the Lagrangian involving members of the ''breve"  family only,
arise from  augmented standard linkages between
\begin{eqnarray}
\label{breve-linkage}
\begin{array}{rcl}
W_\mu^{1j}, W_\mu^{2j}, \;({\breve{{\nu}}}^{e_1}_R, {\breve{e}}_{1R})&\; \mbox{ and }\; &({\breve{{\nu}}}^{e_2}_L, {\breve{e}}_{2L});\\ 
W_\mu^{1j}, W_\mu^{2j}, \;({\breve{u}}_{1R}, {\breve{d}}_{1R})& \;\mbox{ and }\; & ({\breve{u}}_{2L}, {\breve{d}}_{2L}).
\end{array}
\end{eqnarray}

These linkages have the values for the parameters $\xi_1, \; \xi_2 \; \eta_1 \;
\eta_2$ all set equal to +1 as depicted in the diagrammatic representation of the Lagrangian (\ref{Lag}).

The only remaining problem now is to specify the actions of $R_1$ and $R_2$ on
\begin{eqnarray}
\label{list1}
\begin{array}{rcl}
 ({\breve{{\nu}}}^{{e}_1}_L, {\breve{e}}_{1L}, {\breve{\nu}}^{{e}_2}_R,{\breve{e}}_{2R});&&\\
 ({\breve{u}}_{1L}, {\breve{d}}_{1L}, {\breve{u}}_{2R}, {\breve{d}}_{2R}) &&
\end{array}
\end{eqnarray}
and on
\begin{eqnarray}
\label{list2}
\begin{array}{rcl}
({\nu}^{{e}_1}_R, {e}_{1R}, {\nu}^{{e}_2}_L,{e}_{2L});&&\\
({u}_{1R},{d}_{1R}, {u}_{2L}, d_{2L}) &&
\end{array}
\end{eqnarray}
through the appropriate choice of invariants involving the objects listed above (\ref{list1}, \ref{list2})
so that the  three guidelines are fulfilled. This is achieved through the two postulates given below:

$A:$ The masses of the breve electrons and breve quarks are much bigger than the masses of electrons and
quarks respectively.

$B:$ There are augmented Standard Linkage between:
\begin{eqnarray}
\label{list3}
\begin{array}{rcl}
W_\mu^{1j}, W_\mu^{2j}, ({\nu}^{e_1}_R, {\breve{e}}_{1L})\;\;\mbox{and}\;\; ({\nu}^{{e}_2}_L, {\breve{e}}_{2R});&&\\
W_\mu^{1j}, W_\mu^{2j},({\breve{\nu}}^{{e}_1}_L, e_{1R})\;\; \mbox{and}\;\; ({\breve{\nu}}^{{e}_2}_R, {e}_{2L});&&\\
W_\mu^{1j}, W_\mu^{2j}, (u_{1R}, {\breve{d}}_{1L})\;\; \mbox{and} \;\;(u_{2L}, {\breve{d}}_{2R}); &&\\
W_\mu^{1j}, W_\mu^{2j}, ({\breve{u}}_{1L}, d_{1R})\;\; \mbox{and}\;\; ({\breve{u}}_{2R}, d_{2L}).&& 
\end{array}
\end{eqnarray}
with the parameters  ${\xi}_1, {\xi}_2, {\eta}_1, {\eta}_2$ for all these linkages all equal to +1.

Parenthetical remarks: [ Augmented standard linkages $(ASL)$ are described in section (\ref{stage-2}).
Transformations of members of $\mathbf{D} (ASL)$ under $R_1, R_2; U_1, U_1$ are given in sections (\ref{stage-1}); 
(\ref{stage-2}).
respectively. Notice that the transformations under $R_1$, $R_2$ for augmented standard linkages coincide with those for
the standard linkages and are given by
(\ref{variation4}), page (\pageref{variation4}) ]
 
In accordance with expression (\ref{standard-invariant4}), it is now apparent that in the above linkages (\ref{list3}) 
involving object pairs with different handedness the charged weak bosons $(W_{\mu}^{a \pm})$ are absent, in agreement 
with the requirement
of the third guideline. Postulate $A$ further excludes low energy manifestation of couplings of charged
weak bosons to right handed objects which would otherwise have been un-avoidable in view of the postulated
linkages (\ref{breve-linkage}).

The augmented standard invariants appearing in the 
Lagrangian corresponding to the linkages (\ref{list3}) are therefore:

\begin{eqnarray}
\label{standard-invariant3}
\begin{array}{l}
A.H. \left[
i (
\begin{array}{cc}
{\overline{{\nu}^{e_1}}}_R& {\overline{\breve{e}}}_{1L}
\end{array}
)
\gamma^\mu
\{
\partial_\mu
\left\lgroup
\begin{array}{c}
{{\nu}^{e_1}}_R\\
{\breve{e}}_{1L}
\end{array}
\right\rgroup
-
\frac{ig}{2} \sigma^3 W_\mu^{13}
\left\lgroup
\begin{array}{c}
{{\nu}^{e_1}}_R\\
{\breve{e}}_{1L}
\end{array}
\right\rgroup 
- \frac{ig}{2} W_\mu^{23}
\left\lgroup
\begin{array}{c}
{{\nu}^{e_1}}_R\\
{\breve{e}}_{1L}
\end{array}
\right\rgroup
\right. \\[0.15in]
\phantom{A.H. \left[
i (
\begin{array}{cc}
{\overline{{\nu}^{e_2}}}_L& {\overline{\breve{e}}}_{2R}
\end{array}
)
\gamma^\mu
\left\{
\partial_\mu
\left\lgroup
\begin{array}{c}
{{\nu}^{e_2}}_L\\
{\breve{e}}_{2R}
\end{array}
\right\rgroup \right. \right. } 
\left.
+i \frac{f}{2} B^1_\mu  
\left\lgroup
\begin{array}{c}
{{\nu}^{e_1}}_R\\
{\breve{e}}_{1L}
\end{array}
\right\rgroup +
i \frac{f}{2} B^2_\mu 
\left\lgroup
\begin{array}{c}
{{\nu}^{e_1}}_R\\
{\breve{e}}_{1L}
\end{array}
\right\rgroup 
\} \right.\\[0.15 in]
\phantom{
A.H. \left[\right.}
+ i (
\begin{array}{cc}
{\overline{{\nu}^{e_2}}}_L & {\overline{\breve{e}}}_{2R}
\end{array}
)
\gamma^\mu
\{
\partial_\mu
\left\lgroup
\begin{array}{c}
{{\nu}^{e_2}}_L\\
{\breve{e}}_{2R}
\end{array}
\right\rgroup
- \frac{ig}{2} \sigma^3 W_\mu^{13}
\left\lgroup
\begin{array}{c}
{{\nu}^{e_2}}_L\\
{\breve{e}}_{2R}
\end{array}
\right\rgroup  
- \frac{ig}{2} W^{23}_\mu 
\left\lgroup
\begin{array}{c}
{{\nu}^{e_2}}_L\\
{\breve{e}}_{2R}
\end{array}
\right\rgroup\\[.15 in]
\phantom{A.H. \left[
i (
\begin{array}{cc}
{\overline{{\nu}^{e_1}}}_R& {\overline{\mu}}_{2L}
\end{array}
)
\gamma^\mu
\left.\{
\partial_\mu
\left\lgroup
\begin{array}{c}
{{\nu}^{e_1}}_R\\
{\mu}_{2L}
\end{array}
\right\rgroup \right. \right.} 
\left. \left.
-i \frac{f}{2} B^1_\mu  
\left\lgroup
\begin{array}{c}
{{\nu}^{e_2}}_L\\
{\breve{e}}_{2R}
\end{array}
\right\rgroup 
-i \frac{f}{2} B^2_\mu 
\left\lgroup
\begin{array}{c}
{{\nu}^{e_2}}_L\\
{\breve{e}}_{2R}
\end{array}
\right\rgroup 
\} \right]\right.
\end{array} 
\end{eqnarray}

+ terms obtained from previous two through the exchange $(e \;\;\rightleftarrows \breve{e}\;\; \mbox{and} 
\;\;\nu \rightleftarrows \breve{\nu})$ \\
\\
+ terms obtained from previous  four through the substitutions $(e\;\; \rightarrow\;\; d \;\;\mbox{and}\;\;
 {\nu}^e \;\;\rightarrow\;\; u)$ 

\subsection{Discrete Symmetry Transformation (Parity or Mirror Reflection Analogue - MRA)}
\label{parity}

The model developed so far has the following property: for each charged weak interaction
boson ( $W_\mu^{a \pm}, \; \; a=1,2 $) and each object with definite handedness and type ( $e_1, e_2, \nu^{e_1}, \nu^{e_2}, u_1, u_2,
d_1, d_2$ ${\breve{e}}_1, {\breve{e}}_2, {\breve{\nu}}^{e_1}, {\breve{\nu}}^{e_2}, {\breve{u}}_1, {\breve{u}}_2,
{\breve{d}}_1, {\breve{d}}_2$ ), the boson is coupled differently to the right and left handed  
 object of the same type. [ see the terms in the Lagrangian (\ref{Lag}) displayed in  ( \ref{weakterms}  ) ]. There is, however, a
discrete symmetry transformations ( $\Pi$) of the model Lagrangian constructed so far. It is
given by:
\begin{eqnarray}
\label{parity2}
\begin{array}{rcl}
W_\mu^{a \pm} (\vec{x}, t) & \rightarrow & (\delta_\mu^0 - \Sigma_j \delta_\mu^j) \; W_\mu^{a\pm}
(-\vec{x}, t)\\
&&\\
&&\\
{\cal Z}_\mu^a (\vec{x}, t) & \rightarrow & (\delta_\mu^0 - \Sigma_j \delta_\mu^j) \; {\cal Z}_\mu^a
(-\vec{x}, t)\\
&&\\
&&\\ 
{\cal A}_\mu^a (\vec{x}, t) & \rightarrow & (\delta_\mu^0 - \Sigma_j \delta_\mu^j) \; {\cal A}_\mu^a
(-\vec{x}, t)\\

\end{array}
\end{eqnarray}

\begin{eqnarray}
\label{parity1}
\begin{array}{rcl}

                                &&\gamma^0 e_{a \tilde{h}}(-\vec{x} ,t) \mbox{  upper mode}\\
e_{ah}(\vec{x} ,t) & \rightarrow &\\
                                &&\gamma^0 e_{\tilde{a} \tilde{h}}(-\vec{x} ,t) \mbox{  lower mode}\\
&&\\
&&\\
&&\\
                                &&\gamma^0 {\nu_{\tilde{h}}}^{e_{a}}(-\vec{x} ,t) \mbox{   upper mode}   \\
\nu_{h}^{e_a}(\vec{x} ,t) & \rightarrow &\\
                                &&\gamma^0 {\nu_{\tilde{h}}}^{e_{\tilde{a}}}(-\vec{x} ,t) \mbox{   lower mode}\\
&&\\
&&\\
&&\\
                                &&\gamma^0 u_{a \tilde{h}}(-\vec{x} ,t) \mbox{  upper mode}\\
u_{ah}(\vec{x} ,t) & \rightarrow &\\
                                &&\gamma^0 u_{\tilde{a} \tilde{h}}(-\vec{x} ,t) \mbox{  lower mode}\\
&&\\
&&\\
&&\\
                                &&\gamma^0 d_{a \tilde{h}}(-\vec{x} ,t) \mbox{   upper mode}   \\
d_{ah}(\vec{x} ,t) & \rightarrow &\\
                                &&\gamma^0 d_{\tilde{a} \tilde{h}}(-\vec{x} ,t) \mbox{   lower mode}\\
\end{array}
\end{eqnarray}
\\
\\
\\
and transformations obtained from (\ref{parity1}) through the substitutions $e \rightarrow \breve{e},\;\; \nu \rightarrow 
\breve{\nu},\\
u \rightarrow \breve{u}, \;\;d \rightarrow \breve{d}$
\\

where $ a\; =\; 1\;,\; 2\; ; \;h\; =\; L\;,\; R\; ;\; \tilde{1}\; = \;2\;,\; \tilde{2}\; =\; 1\;;\;
 \tilde{L}\; = \;R\;,\; \tilde{R} \;=\; L $ and $(\vec{x}, t)$ are the space and time components of space-time
co-ordinates x. The factors $(\delta_\mu^0 - \Sigma_j \delta_\mu^j)$, on the right of arrows, above have the effect of 
making only the space components of the gauge fields to change sign on the right hand side. The upper [lower] modes
of operation for the transformation is manifested in those terms ( i.e. monomials) in the Lagrangian in which 
 ${\cal{A}}_\mu^a\;, \;{\cal{Z}}_\mu^a$ [$ W_\mu^a$] occurs.

The discrete symmetry transformation described above (probably) can not be represented by the action of some 
operator acting on the (Hilbert) space of particle states in the model. This is quite unlike the standard model 
where a discrete symmetry somewhat analogous to the discrete symmetry of the Lagrangian described here is
represented by an operator $P$ in the state space. In the context of the standard model it is then meaningful
to investigate $P, CP, PT, CPT$ (non) conservation. Such questions do not seem to be well-defined in the present model.
Notice, however, that the discrete transformation above applied twice is equivalent to the identity.
\section{Lagrangian of the model}
\label{lagrange}

We shall present diagrammatic representation of the part of the model Lagrangian
which does not involve the Higgs scalars \cite{Higgs1, Higgs2, Higgs3, Ander1, Brout} and the associated mass terms -
i.e. we shall restrict ourselves to presenting explicitly only the left hand end of the complete Lagrangian. We shall
consider the first family of leptons and quarks only.
The right hand end of the Lagrangian and the associated new symmetries linking scalars, spinors and vectors is a separate 
topic for future presentation. Using the notation for augmented standard invariants described on page (\pageref{stage-2}),
the Lagrangian of the model can now be written as
\begin{eqnarray*}
\label{Lag}
\begin{array}{l}
W^{2}_{\mu}[1,1] \left[
\begin{array}{c}
W^{1}_{\mu}[1,1] \left\{
\begin{array}{c}
\begin{array}{ccc}
{{\nu}^{e_1}}_L & -\frac{1}{2} & -\frac{1}{2}
\end{array}\\
\\
\begin{array}{ccc}
e_{1L}\; & \;-\frac{1}{2} & -\frac{1}{2}
\end{array}
\end{array}
\right.\\
\\
W^{1}_{\mu}[1,1]\left\{
\begin{array}{c}
\begin{array}{ccc}
{{\nu}^{e_2}}_R & \;\;\frac{1}{2} & \;\;\;\frac{1}{2}
\end{array}\\
\\
\begin{array}{ccc}
e_{2R}\;\; & \;\;\frac{1}{2} & \;\;\;\frac{1}{2}
\end{array}
\end{array}
\right.
\end{array}
\right.\\
\\
+W^{2}_{\mu}[1,1] \left[
\begin{array}{c}
W^{1}_{\mu}[1,1] \left\{
\begin{array}{c}
\begin{array}{ccc}
{{\breve{\nu}}^{e_1}}_R & -\frac{1}{2} & -\frac{1}{2}
\end{array}\\
\\
\begin{array}{ccc}
{\breve{e}}_{1R}\; \;& -\frac{1}{2} & -\frac{1}{2}
\end{array}
\end{array}
\right.\\
\\
W^{1}_{\mu}[1,1]\left\{
\begin{array}{c}
\begin{array}{ccc}
{{\breve{\nu}}^{e_2}}_L &\;\; \frac{1}{2} &\;\;\; \frac{1}{2}
\end{array}\\
\\
\begin{array}{ccc}
{\breve{e}}_{2L}\;\; & \; \;\frac{1}{2} & \;\;\; \frac{1}{2}
\end{array}
\end{array}
\right.
\end{array}
\right.
\end{array}
\end{eqnarray*}
\begin{eqnarray*}
\begin{array}{l}
+W^{2}_{\mu}[1,1] \left[
\begin{array}{c}
W^{1}_{\mu}[1,1] \left\{
\begin{array}{c}
\begin{array}{ccc}
{{\nu}^{e_1}}_R & -\frac{1}{2} & -\frac{1}{2}
\end{array}\\
\\
\begin{array}{ccc}
{\breve{e}}_{1L}\; & \;-\frac{1}{2} & -\frac{1}{2}
\end{array}
\end{array}
\right.\\
\\
W^{1}_{\mu}[1,1]\left\{
\begin{array}{c}
\begin{array}{ccc}
{{\nu}^{e_2}}_L & \;\;\frac{1}{2} & \;\;\;\frac{1}{2}
\end{array}\\
\\
\begin{array}{ccc}
{\breve{e}}_{2R}\;\; & \;\;\frac{1}{2} & \;\;\;\frac{1}{2}
\end{array}
\end{array}
\right.
\end{array}
\right.\\
\\
+W^{2}_{\mu}[1,1] \left[
\begin{array}{c} 
W^{1}_{\mu}[1,1] \left\{
\begin{array}{c}
\begin{array}{ccc}
{{\breve{\nu}}^{e_1}}_L & -\frac{1}{2} & -\frac{1}{2}
\end{array}\\
\\
\begin{array}{ccc}
e_{1R}\; \;& -\frac{1}{2} & -\frac{1}{2}
\end{array}
\end{array}
\right.\\
\\
W^{1}_{\mu}[1,1]\left\{
\begin{array}{c}
\begin{array}{ccc}
{{\breve{\nu}}^{e_2}}_R &\;\; \frac{1}{2} &\;\;\; \frac{1}{2}
\end{array}\\
\\
\begin{array}{ccc}
e_{2L}\;\; & \; \;\frac{1}{2} & \;\;\; \frac{1}{2}
\end{array}
\end{array}
\right.
\end{array}
\right.
\end{array}
\end{eqnarray*}
\begin{eqnarray}
\begin{array}{l}
+ W^{2}_{\mu}[1,1] \left[
\begin{array}{c}
W^{1}_{\mu}[1,1] \left\{
\begin{array}{c}
\begin{array}{ccc}
u_{1L} & -\frac{1}{6} & -\frac{1}{6}
\end{array}\\
\\
\begin{array}{ccc}
d_{1L}\ & -\frac{1}{6} & -\frac{1}{6}
\end{array}
\end{array}
\right.\\
\\
W^{1}_{\mu}[1,1]\left\{
\begin{array}{c}
\begin{array}{ccc}
u_{2R} & \;\;\frac{1}{6} & \;\;\;\frac{1}{6}
\end{array}\\
\\
\begin{array}{ccc}
d_{2R}\;\; & \;\;\frac{1}{6} & \;\;\;\frac{1}{6}
\end{array}
\end{array}
\right.
\end{array}
\right.\\
\\
\begin{array}{c}
+ W^{2}_{\mu}[1,1] \left[
\begin{array}{c}
W^{1}_{\mu}[1,1] \left\{
\begin{array}{c}
\begin{array}{ccc}
{\breve{u}}_{1R} & -\frac{1}{6} & -\frac{1}{6}
\end{array}\\
\\
\begin{array}{ccc}
{\breve{d}}_{1R}\ & -\frac{1}{6} & -\frac{1}{6}
\end{array}
\end{array}
\right.\\
\\
W^{1}_{\mu}[1,1]\left\{
\begin{array}{c}
\begin{array}{ccc}
{\breve{u}}_{2L} & \;\;\frac{1}{6} & \;\;\;\frac{1}{6}
\end{array}\\
\\
\begin{array}{ccc}
{\breve{d}}_{2L}\;\; & \;\;\frac{1}{6} & \;\;\;\frac{1}{6}
\end{array}
\end{array}
\right.
\end{array}
\right.
\end{array}\\
\\
\end{array}
\end{eqnarray}
\begin{eqnarray*}
\begin{array}{l}
+ W^{2}_{\mu}[1,1] \left[
\begin{array}{c}
W^{1}_{\mu}[1,1] \left\{
\begin{array}{c}
\begin{array}{ccc}
{\breve{u}}_{1L} & -\frac{1}{6} & -\frac{1}{6}
\end{array}\\
\\
\begin{array}{ccc}
d_{1R}\ & -\frac{1}{6} & -\frac{1}{6}
\end{array}
\end{array}
\right.\\
\\
W^{1}_{\mu}[1,1]\left\{
\begin{array}{c}
\begin{array}{ccc}
{\breve{u}}_{2R} & \;\;\frac{1}{6} & \;\;\;\frac{1}{6}
\end{array}\\
\\
\begin{array}{ccc}
d_{2L}\;\; & \;\;\frac{1}{6} & \;\;\;\frac{1}{6}
\end{array}
\end{array}
\right.
\end{array}
\right.\\
\\
\begin{array}{c}
+ W^{2}_{\mu}[1,1] \left[
\begin{array}{c}
W^{1}_{\mu}[1,1] \left\{
\begin{array}{c}
\begin{array}{ccc}
u_{1R} & -\frac{1}{6} & -\frac{1}{6}
\end{array}\\
\\
\begin{array}{ccc}
{\breve{d}}_{1L}\ & -\frac{1}{6} & -\frac{1}{6}
\end{array}
\end{array}
\right.\\
\\
W^{1}_{\mu}[1,1]\left\{
\begin{array}{c}
\begin{array}{ccc}
u_{2L} & \;\;\frac{1}{6} & \;\;\;\frac{1}{6}
\end{array}\\
\\
\begin{array}{ccc}
{\breve{d}}_{2R}\;\; & \;\;\frac{1}{6} & \;\;\;\frac{1}{6}
\end{array}
\end{array}
\right.
\end{array}
\right.
\end{array}\\
\\
+\;\mbox{invariants of gauge fields only} \\
\\
+ \mbox{invariants involving Higgs scalars}. 
\end{array}
\end{eqnarray*}

\section{Possible Consequences of the Results of the Model for Nuclei (and Chemistry) }
\label{nuclei}

This section deals mainly with exploring the consequences of the two types of protons 
and neutrons for nuclei. However we shall also briefly mention some consequences for chemistry to the 
extent that it would be be helpful in clarifying the associated ideas.

We proceed by relying on a plausible heuristic model of nuclei composed of the
two types of nucleons. The emphasis would be on presenting a method for deriving empirical relations for the masses
of ''isomers" (i.e. nuclides with a definite atomic number $Z$ and mass number $A$, but with
different detailed constitutions determined by number of nucleons of each type). These mass
differences are quite small and are on the borderline of limits on experimental precision in measurements
of nuclear isotope masses \cite{Isotopes}.
 
Each nucleus of atomic number $Z$ may have $Z_1$ protons ($p$) and $Z_2$ p-protons ($p'$)
where
\begin{eqnarray}
Z & = & Z_1 + Z_2
\end{eqnarray}

Parenthetical remark: [ the notation ''p-proton" stands for primed proton ($p'$) and ''u-neutron" stands for
unprimed neutron ($n$),  proton ($p$) is unprimed and neutron ($n'$) is primed.]

If the same nucleus has atomic mass number $A$ then it may have $N_1$ u-neutrons ($n$) and
$N_2$ neutrons ($n'$)
where
\begin{eqnarray}
A-Z & = & N_1 + N_2
\end{eqnarray}

It can be shown that the total number of isomers
of a nuclide ($Z, A$) is
\begin{eqnarray}
\label{relation1}
N_{isom.} & = & (A-Z+1)(Z+1)
\end{eqnarray}

These isomers would be of nearly identical masses only if $p,\; p',\; n,\; n'$ are all nearly equal in mass which
implies that  $m(u_1),\; m(d_1),\;m(u_2),\;m(d_2)$ have nearly equal masses.
The discussion in this paper is restricted to consideration of this possibility only. Scenario $Y$ is not 
being considered here.

Thus, for example Triton ( $A=3,\; Z=1$ ) has the six isomers shown in Figure (7) when viewed in ${\cal{ A}}^1_\mu$.
However these same nuclides when viewed in $ {\cal {A}}^2_\mu $ would appear as a mixture of isotopes of several chemically
different elements.

\begin{figure}[htbp]
\centering{\mbox{\psfig{figure=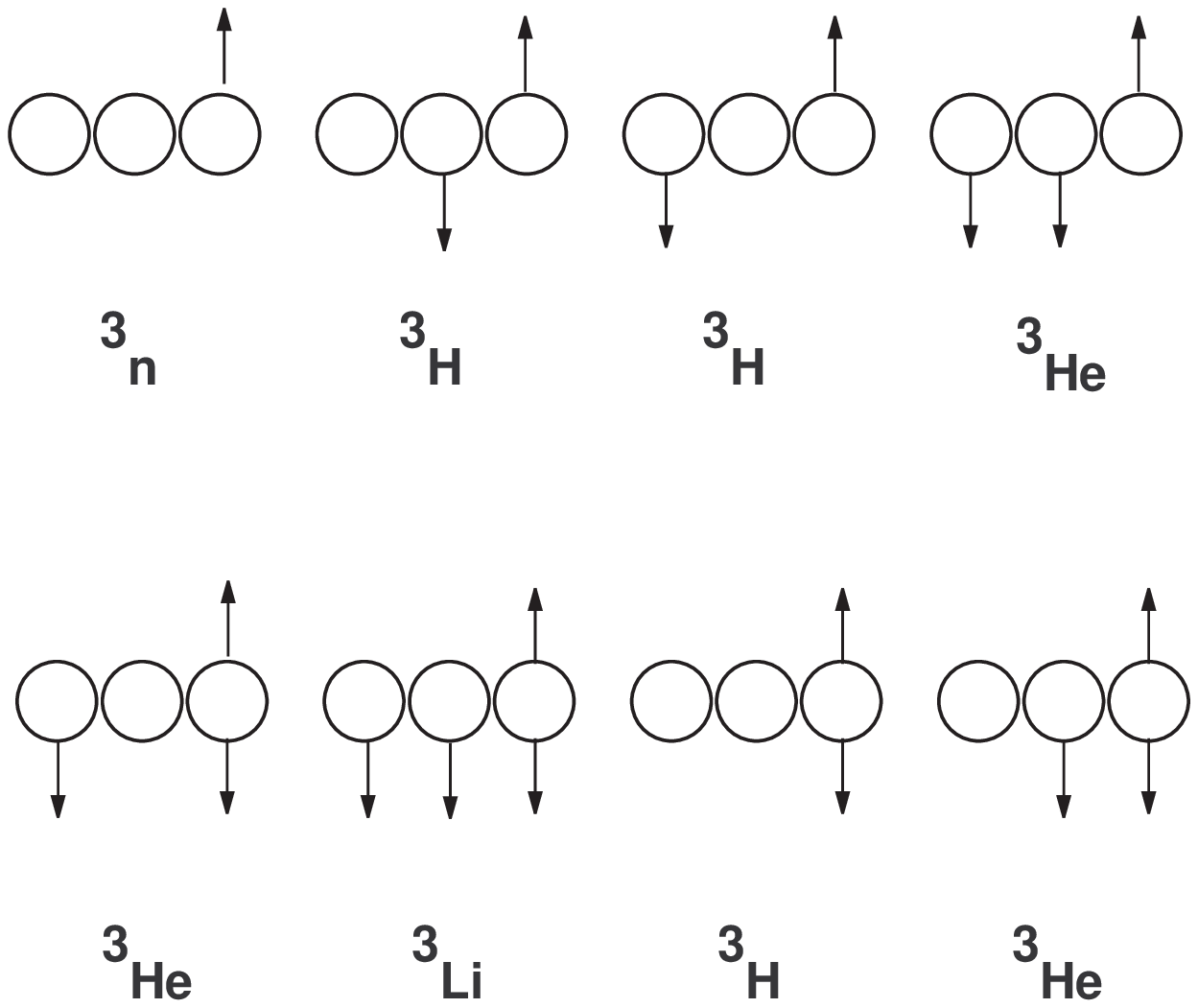}}}
\label{tritium-fig}
\caption{Triton in ${\cal{A}}_\mu^1$ appearing as a mixture in ${\cal{A}}_\mu^2$}
\end{figure}

Introduce the symbols $\circle$ and $\down$ for the two kinds of neutrons $(n', n)$, with
$Q_1$ values $(0,0)$ and $Q_2$ values $(0,1)$ respectively; the symbols
$\up$ and $\both$ for the two kinds of protons $(p',p)$, with $Q_1$ values
$(1,1)$ and $Q_2$ values $(0,1)$ respectively.  Then the potential isomers for
$^3H_1$ (Triton), possibly unstable, may be represented as shown in Figure (7).

[Notice that the middle two configurations in the top row of Figure (7) are identical and the two end configurations in the
bottom row are identical, so that there are six distinct configurations (in agreement with relation (\ref{relation1}).]

These isomers could capture any free ${\overline{e}}_2$'s in their neighborhood. Since the ${\overline{e}}_2$'s 
(attached to the isomers) are coupled to $ {\cal{A}}^2_\mu $ they would provide the foundation for a parallel 
chemistry involving the same nuclides. This parallel or alternative chemistry (''al-chemistry" ?) would then be 
present in our world but its operation may not be amenable to the usual methods of chemical analysis as would become
apparent from the description of scenario $S_1$ presented in section (\ref{sec6}), page (\pageref{sec6}).

A collection of neutral atoms ($Q_1 =0, \; Q_2=0$) with  nuclei of atomic number $Z$ $(\mbox{i.e.}\; Q_1\; = \;Z,\; Q_2
\mbox{ un-defined})$
and mass number $A$ [in  $(A-Z+1)(Z+1)$ distinct isomeric states] when illumined with light of the first 
kind $({\cal{A}}_\mu^1)$
would appear as a chemically pure substance as far as chemical reactions involving $e_1$ and ${\cal{A}}_\mu^1$ go. However, 
when viewed with light of the second kind $({\cal{A}}_\mu^2)$ the collection is no longer chemically pure from the
perspective of chemical reactions involving $\overline{e}_2$ and ${\cal{A}}_\mu^2$. It is then a mixture of isotopes of 
different chemical elements. 

Thus, in  Figure (7) the various isomers of $^3H_1$ (only six of these are distinct) will 
appear as $^3n$, $^3H_1$,$^3H_1$, $^3He_2$,$^3He_2$,$^3Li_3$, $^3H_1$, and $^3He_2$.

We shall now present an outline of the method for determining the approximate masses of these isomers of 
nuclide ($Z, A$) in terms of the known masses of the nuclides  ($Z, A$) given in the Tables of Isotopes \cite{Isotopes}. 
In this calculation the masses of $p, p', n, n'$ appear as parameters whose best values are determined by invoking the 
following hypothesis:

$H_1$: The masses in the Tables of isotopes are those for isomers with lowest mass.

Consider a nucleus with atomic number $Z$ and mass number $A$.  Let us try to determine the mass of the nucleus
in an isomeric state in which it has $Z_1$ protons $(p)$ of mass $m(p)$) and $Z_2$ p-protons of mass $m(p')$, $N_1$
u-neutrons $(n)$ of mass $m(n)$ and $N_2$ neutrons $(n')$ of mass $m(n')$

The total mass of the system, for large atomic mass number $A$, may be approximated as

\begin{eqnarray}
\label{systemmass}
\begin{array}{ccl}
M(Z,A;Z_1,N_1) =  Z_1 m(p) + (Z-Z_1) m(p')&&\\[0.05 in]
\;\;\;\;+ N_1 m(n) +(A-Z-N_1) m(n') && \\[0.05 in]
\;\;\;\;+ \frac{1}{c^2}[\mbox{ Nuclear binding energy }&&\\ [0.05 in]
\;\;\;\; + \mbox{ Coulomb repulsion energy }&&\\[0.05 in]
\;\;\;\; + \mbox{Kinetic Energies of Fermi gases}]&&
\end{array}
\end{eqnarray}

where
\begin{eqnarray*}
\begin{array}{ccl}
\mbox{Nuclear binding energy} = \frac{1}{2}n_b\;b\;A &&\\[0.1 in]
\mbox{Coulomb repulsion energy} = [Z(Z-1)+&&\\[0.05 in]
\;\;\;\;(N_1 + Z_1)(N_1+Z_1-1)]\frac{g^2 \xi}{2a}&&\\[0.1 in]
\mbox{Kinetic energies} = \frac{\lambda c^2}{a^2}(Z_1^{\frac{5}{3}} \frac{1}{m(p)} &&\\[0.05 in]
\;\;\;\;+ (Z-Z_1)^{\frac{5}{3}} \frac{1}{m(p')} + N_1^{\frac{5}{3}} \frac{1}{m(n)} &&\\[0.05 in]
\;\;\;\;+ (A-Z-N_1)^{\frac{5}{3}} \frac{1}{m(n')})&&\\[0.2 in]
\lambda = \frac{3}{10} (\frac{9 \pi}{4})^{\frac{2}{3}}\frac {\hbar^2}{c^2}&&
\end{array}
\end{eqnarray*}

$n_b$ is the average number of nearest neighbors of a nucleon and $-b$ is the nuclear binding energy per nuclear bond
(assumed to occur between pairs of nearest neighbors only).  $a$ is the radius of the spherical region over which the 
charges carried by the nucleons are confined. $\xi$ is defined such that for a sphere of radius $a$ the average value 
of $\frac{1}{r}$ between two points taken at random is $\frac{\xi}{a}$:  $\;\xi = \frac{6}{5}$.

Since we are making the simplifying assumption that the binding energy $-b$ for $pp, pp',pn, $$pn', nn, np',$ $nn', p'p', p'n', n'n'$ bonds are the same, we expect the experimental 
data on nuclear isotopes to exhibit an apparently random jitter around the masses derived in the model.

If it is assumed that $a$ depends on $A$ only then the minimum mass isomer has values for $Z_1, N_1$ given by
the integers nearest in value to the solutions of the equations

\begin{eqnarray}
\begin{array}{lcl}
m(p)-m(p') + (2Z_1 + 2N_1 -1)\frac{g^2 \xi}{2a} +&&\\[0.05 in]
\;\;\;\; \frac{5\lambda}{3a^2}(Z_1^{\frac{2}{3}}\frac{1}{m(p)} -(Z-Z_1)^{\frac{2}{3}}\frac{1}{m(p')})=0 &&\\[0.1 in]
 m(n)-m(n') + (2Z_1 + 2N_1 -) \frac{g^2 \xi}{2a} + &&\\[0.05 in]
\;\;\;\; \frac{5\lambda}{3a^2} (N_1^{\frac{2}{3}}\frac{1}{m(n)} - (A-Z-N_1)^{\frac{2}{3}}\frac{1}{m(n')})=0&&
\end{array} 
\end{eqnarray}

Substitution of the values for $Z_1, N_1$ in (\ref{systemmass}) then gives the mass $M(A,Z)$ of the isomer with lowest mass.

The mass increments derived for the other isomers are thus found to be within the limits of experimental uncertainties in 
the tables of isotopes \cite{Isotopes}.

We now present an analysis of the two scenarios $( X_1, X_2) $ mentioned in section (\ref{sec2.2})  
\label{scenx}

Scenario $X_1$: In this scenario we have $m(u_1) = m(u_2),\; \; m(d_1)=m(d_2)$ and a constraint on
$f, g $ and the independent mass ratios  among $m(e),\; m(u),\; m(d),\;$
$m(W)\; \mbox{and} \; m(Z)$. This leads to $ m(p) = m(n'),\; \; m(p') = m(n)$. 
\\
\\
\\
Two sub-scenarios $(X_{11}, X_{12})$ arise
\\
\\
\\
$X_{11}$: The proton $p$ is the observed stable proton. Then the u-neutron $n$ must be the observed 
unstable neutron (which has mass different from that of the proton $p$). In this sub-scenario the p-proton $p'$ 
(with mass equal to that of the u-neutron $n$) decays with a lifetime equal to that of the u-neutron $n$ according to 
the following transitions. 
\begin{eqnarray}
\label{decay1}
\begin{array}{rcl}
p' & \rightarrow & p + {\overline{e}}_1 + \overline{{\nu}^{e_2}}\\
p' & \rightarrow & n' + {\overline{e}}_1 + {\nu}^{e_1}\\
p' & \rightarrow & p + {\overline{e}}_2 + \overline{{\nu}^{e_1}}\\
p' & \rightarrow & n' + {\overline{e}}_2 + {\nu}^{e_2}
\end{array}
&&
\end{eqnarray}

Notice that the first two decays mimic some aspects of the decay of an anti-neutron (in the standard model)
Also in this scenario the u-neutron $n$ decays according to
\begin{eqnarray}
\label{decay5}
\begin{array}{rcl}
n & \rightarrow & p +  e_1 + \overline{{\nu}^{e_1}}\\
n & \rightarrow & n' + e_1 + {\nu}^{e_2}\\
n & \rightarrow & p + e_2 + \overline{{\nu}^{e_2}}\\
n & \rightarrow & n' + e_2 + {\nu}^{e_1}
\end{array} 
&&
\end{eqnarray}
 
In this scenario the observed part of the solar neutrino flux, being measured by experiments designed to look for a
(binary) inverse of the first reaction in (\ref{decay5}) only , is thus expected to be one fourth of the result derived 
from the standard model (plus the solar model \cite{Adelberger}) if  the modifications introduced by the present model 
into the solar model itself are ignored. This fact would seem to favor this scenario or the next one 
as most likely to have been realized in nature.

$X_{12}$: The p-proton $p'$ is the observed stable proton. Then the neutron $n'$ must be the observed unstable neutron
which has mass different from that of $p'$. In this sub-scenario the proton $p$ (with mass equal to that
of the neutron $n'$) decays with a lifetime equal to that of the neutron ($n'$) according to the following
transitions.
\begin{eqnarray}
\label{decay2}
\begin{array}{rcl}
p &\rightarrow & p' + e_1 + {\nu}^{e_2}\\
p &\rightarrow & n + {\overline{e}}_1 + {\nu}^{e_1}\\
p &\rightarrow & p' + e_2 + {\nu}^{e_1}\\
p &\rightarrow & n + {\overline{e}}_2 + {\nu}^{e_2}
\end{array}
&&
\end{eqnarray}

Notice that the first two decays mimic some aspects of the decays of a neutron and an anti-neutron respectively
(in the standard model). Also in this scenario the neutron $n'$ decays according to
\begin{eqnarray}
\label{decay6}
\begin{array}{rcl}
n' & \rightarrow & p' +  e_1 + \overline{{\nu}^{e_1}}\\
n' & \rightarrow & n + {\overline{e}}_2 + \overline{{\nu}^{e_1}}\\
n' & \rightarrow & p' + e_2 + \overline{{\nu}^{e_2}}\\
n' & \rightarrow & n + {\overline{e}}_1 + \overline{{\nu}^{e_2}}
\end{array}
&&
\end{eqnarray}

In this scenario also the observed part of the solar neutrino flux, being measured by experiments designed to look for a
(binary) inverse of the first reaction in (\ref{decay6}) only , is thus expected to be one fourth of the result derived 
from the standard model (plus the solar model \cite{Adelberger}) if  the modifications introduced by the present model 
into the solar model itself are ignored. This fact would seem to favor this  scenario or the previous one 
as most likely to have been realized.

Scenario $X_2$: In this scenario we have $m(u_1) = m(d_2),\; \; m(d_1)=m(u_2)$ and a constraint on
$f, g $ and the independent mass ratios  among $m(e),\; m(u),\; m(d),\;$
$m(W)\; \mbox{and} \; m(Z)$. This leads to $ m(p) = m(p'),\; \; m(n) = m(n')$. 

In this scenario the observed neutron is a mixture of $n$ and $n'$ and decays according to the transitions

\begin{eqnarray}
\label{decay3}
\begin{array}{rcl}
n &\rightarrow &p + e_1 + \overline{{\nu}^{e_1}}\\
n &\rightarrow &p + e_2 + \overline{{\nu}^{e_2}}\\ 
n' &\rightarrow &p' + e_1 + \overline{{\nu}^{e_1}}\\
n' &\rightarrow &p' + e_2 + \overline{{\nu}^{e_2}}
\end{array}
\end{eqnarray}

Notice that in this scenario each type of neutron has only two allowed kinematically equivalent decay channels. Contrasting
this with the following four kinematically equivalent decay channels for the decay of $ \mu_1 $ ( which is the analogue
of $\mu^-$ in the standard model)
\begin{eqnarray}
\label{decay4}
\begin{array}{rcl}
\mu_1 &\rightarrow & e_1 + {\nu}^{{\mu}_1}+ \overline{{\nu}^{e_1}}\\
\mu_1 &\rightarrow & e_1 + \overline{{\nu}^{{\mu}_2}} + {\nu}^{e_2}\\
\mu_1 &\rightarrow & e_2 + {\nu}^{{\mu}_1}+ \overline{{\nu}^{e_2}}\\
\mu_1 &\rightarrow & e_2+ \overline{{\nu}^{{\mu}_2}} + {\nu}^{e_1}
\end{array}
\end{eqnarray}

we see that the experimentally observed near equality of phenomenological Fermi coupling constants describing muon and beta
decay \cite{Riazuddin} rules out this scenario in the present model with $M_{W^1} = M_{W^2},$  $g_1= g_2$ . However this 
scenario may be resurrected if one goes beyond the present model and imposes the condition
\\
\\
\\
\begin{eqnarray*}
\frac {g_2^2} {M_{W^2}^2} & << & \frac{g_1^2} { M_{W^1}^2}
\end{eqnarray*}
\\
where $g_1, g_2$ are the coupling constant associated with $R_1, R_2$ and $M_{W^1}, M_{W^2}$ are the masses of the
charged weak bosons.

Scenario $X_3$: In this scenario it is postulated that

\begin{eqnarray}
\label{condition}
 N_1 \;=\;0  & ,\; & Z_1\;=\;0 
\end{eqnarray}

for all nuclei in their ground states. 
Thus all nuclei in their ground state have $Q_2 =0$ and they do not capture any ${\overline{e}}_2$'s in their vicinity.
For such nuclei in their ground state there is one chemistry based on $e_1$'s and ${\cal{A}}_{\mu}^1$  and the other
chemistry based on $e_2$'s and ${\cal{A}}_{\mu}^2$ can not occur. Since this condition (\ref{condition}) does not
arise for the ground state in the heuristic model for nuclei based on hypothesis $H_1$, proposed above, we can say that 
for this scenario to occur we may have either

Scenario $X_{31}$: 

In this scenario the following hypothesis holds true.

$H_2$: It is possible to include an additional term in the model for nuclei which has the desired effect of
producing the condition (\ref{condition}) without producing results in violation of the empirical data on
nuclear masses.

or,

Scenario $X_{32}$:   
In this scenario the following hypothesis is assumed to hold:

$H_3$: It has been arranged by the inter-dependent web of life and environment in our vicinity 
(GAIA \cite{Lovelock, Lovelock2}) that the nuclei with $N_1=0$ and $Z_1=0$ have been retained and
all other nuclei excluded during the course of a few billion years of evolving life forms.

Scenario $X_4$: This scenario envisages the possibility of the two parallel chemistries being
identical and producing effects merely reinforcing each other and not causing matter to evolve
along competing tracks of chemical evolution. It is suggested that the following paragraph be read
immediately after the next section (\ref{sec6}) when the reader would have become familiar
with the notion of the four types of matter.

In this scenario, we envisage that the lowest mass isomers of nuclide $A,Z$ are always such that $Z_2$ 
(the number of p-protons, $p'$) is equal to $N_1$ (the number of u-neutrons $n$). This has the consequence that, in this
scenario, the chemical identity of an atom of Type III matter with any nucleus (in its nuclear ground state)
is the same in the two parallel chemistries - i.e. one chemistry that is based on $e_1$ and ${\cal{A}}^1$
and the other chemistry that is based on ${\overline{e}}_2$ and ${\cal{A}}^2$. This is because
the atomic number relevant to the first chemistry being $Z_1 + Z_2$ is equal in this scenario to the 
atomic number relevant to the second chemistry which is $Z_1 + N_1$. Thus in this scenario the two 
observers depicted in figure (1) when looking at an atom of Type III matter, with its nucleus in the
state of lowest energy, would see the same chemical object.

This scenario gives rise to sub-scenarios $X_{41}$ and $X_{42}$ analogously to $X_{31}$ and $X_{32}$, according
as hypotheses analogous to $H_2$ or $H_3$ hold.

\section{Four Types of Matter and Anti-matter and Three Scenarios of Physical Theory}
\label{sec6}
We have seen that in this model baryons and anti-baryons have
$Q_1, Q_2 > 0$ and $Q_1, Q_2 < 0$ respectively.  Four broad types of matter (anti-matter) are
therefore possible.

Type I: Nuclei made of baryons (anti-baryons) enclosed in shells of
$e_1$s ($\overline{e_1}$s) neutralizing the $Q_1$ charge on the nuclei
while the $Q_2$ charge is not neutralized.

Type II: Nuclei made of baryons (anti-baryons) enclosed in shells of
$\overline{e_2}$s ($e_2$s) neutralizing the $Q_2$ charge on the nuclei
while the $Q_1$ charge is not neutralized.

Type III: Nuclei made of baryons (anti-baryons) enclosed in shells of
$\overline{e_2}$s ($e_2$s) and $e_1$s ($\overline{e_1}$) neutralizing
both kinds of charges on the nuclei.

Type IV: Nuclei made of baryons (anti-baryons) with both kinds of charge
active.

Thus in a bath of electromagnetic radiation of the first ( second) kind [ i.e.,
${\cal{A}}_\mu^1( {\cal{A}}_\mu^2)]$
matter of Type III would lose $e_1$ ($e_2$) and get converted into
matter of the type II ( I ). Notice, also that nuclei of matter of the first  type
have a barrier against annihilation with nuclei of anti-matter of the second  type.
This is so because the {\em shells} of $e_1$ having $Q_1 < 0\; \mbox{and}\; Q_2 = 0$
{\em surrounding nuclei} of matter of the first type  do not 
directly interact with {\em shells} of $e_2$ having
$Q_2 > 0 \;\mbox{and}\; Q_1 = 0 $
{\em surrounding nuclei} of antimatter of the second type. So these
shells are, respectively, exposed to direct interaction with the nuclei of  anti-matter of the
the second type and nuclei of matter of the first type (i.e. the non-corresponding nuclei). This direct 
interaction between the shells and non-corresponding nuclei is therefore repulsive and provides a barrier
against mutual annihilation of the nuclei. However, so long as aggregates of the two different types 
of atoms are far apart, there is no long range (varying as $\frac{1}{r^2}\; \mbox{or as}\; \frac{1}{r^3}$ ) 
force, except for gravitation, between them. 

One might be inclined to think that the type III  matter is the most common form of matter in
our solar system since there are no strong long-range forces other
than the weak force of gravity among the sun and the planets.
Matter of Types I (II) would seem to be precluded from occurrence on large 
scale in the solar system since the strong repulsive
force due to charge $Q_1$ ($Q_2$) carried by the two types of matter
would necessitate the existence of  an attractive force much stronger than the weak 
attractive force of gravity derived from the phenomenology of the solar system.
On the other hand, we have to bear in mind the fact that this extremely weak force
of gravity is believed to be a fundamental force of nature by the
consensus of opinion prevalent at present. 

So we can now argue that in order for this model to make empirical sense,
in the context of the prevalent notion about the nature of weak gravity 
as a fundamental force, one of the following scenarios may hold

scenario $S_1$: matter on earth and in the solar system is almost entirely of Type III .
If this is so then matter experienced by us is flooded with light of both kinds
arising from atomic transitions involving both types of electrons. However as
we have seen above a nuclide with a definite value of nuclear charge $Q_1$ (
manifested when interacting with light ${\cal{A}}_\mu^1$) may have one of several
possible values of the nuclear charge $Q_2$ (manifested when interacting with light 
${\cal{A}}_\mu^2$).

So in the case of matter of Type III we would see the spectral lines due to the second kind
of light produced in transitions of $e_2$'s among the energy levels of atoms with (several) corresponding
chemical alternatives ($Q_2$ values) even if we were looking at materials of high chemical purity (definite
$Q_1$ values) which we can easily prepare with the tools at our disposal. But this is clearly
not observed to be the case.

Therefore we conclude that either of the following two sub-scenarios may occur.

sub-scenario $S_{11}$: matter accessible to us on earth, at least, is of Type I and not Type III. It might 
have been created from a primordial Type III state by a floodlight of the second kind which swept away most
of the $e_2$'s (Some of the residual $e_2$'s left over since the primeval floodlight are manifested in
the phenomena of superconductivity - see section (\ref{supercond}), page (\pageref{supercond}).
In this sub-scenario if, in agreement with the consensus, 
the role of the weak gravity as a fundamental force of nature is to be maintained then
we can not allow the strong long range forces (due to the two kinds of charges) to operate and therefore
apart from the earth (which is made of Type I matter) all the planets and the sun are composed of
Type III matter. In other words, we have to give up the Copernican doctorine that the earth is not special
in any fundamental way relative to the rest of the solar system.

sub-scenario $S_{12}$: matter accessible to us on earth is of Type III (as in the rest of the solar system). However
we are not seeing the spectral lines due to the second kind of light because our consciousness is locked
on to the chemical reactions in the retina involving $e_1$'s and ${\cal{A}}_\mu^1$ and is oblivious to
the chemical reactions in the retina involving $\overline{e}_2$'s and ${\cal{A}}_\mu^2$. This would explain the non-
observation of the spectral line due to the second kind of light produced in transitions of $\overline{e}_2$'s in 
atoms of matter of Type III. Also notice that our photographic plates have been designed by us to be attuned to
light of the first kind. Since the nuclides in the photographic plate with a definite value of $Q_1$ correspond
 to several different values of $Q_2$ we do not expect the Type III atoms associated with these nuclides to
be also attuned to a chemical process triggered by light of the second kind. Thus our present day photographic plates are
not constructed to be sensitive to light of the second kind and we would not see the imprint of the spectral lines
due to light of the second kind on these plates. 
 
Finally, we consider the scenario of rejecting the prevailing consensus about the nature of weak gravity as a fundamental
force.

scenario $S_2$: matter on earth, the planets and the sun is of the same  Type ( I ), in conformity with the Copernican 
doctorine. There are strong repulsive forces between the sun, earth and planets due to the $Q_2$ charges carried 
by matter of Type I. These repulsive forces may be countered by almost equally strong gravitational forces,
resulting in a net weak
attractive force of classical gravity, in order
to explain the phenomenology of the solar
system. However, in order to realize this scheme in a natural fashion we can not have the residual
weak gravity as a fundamental force. This is because the $Q_2$ charges carried by earth, the sun and the planets would
depend on the temperatures of these bodies. The repulsive forces in the solar systems are therefore dependent
on the temperatures. The strong attractive gravitational force, being a fundamental force, is independent of temperatures.
The residual force of weak gravity is therefore dependent on the temperatures and is not a fundamental force.
This scenario could therefore be tested in an accurate version of the Cavendish experiment in which the gravitating bodies
are maintained at different variable temperatures. 
Notice that in this scenario the fundamental force of gravity is almost as strong as electromagnetism of either kind 
and therefore one expects the problem of unification of gravity with all the forces to be simpler. Also, notice that in 
this scenario the residual force of weak gravity holding the sun and planets together would increase (decrease) with decreasing 
(increasing) temperature of the sun (since the charge $Q_2$ producing repulsive forces is expected to decrease (increase)
with decreasing (increasing) temperature).
This has the consequence of making the solar system a far more hospitable place for the evolution and continuation of life
since there is a mechanism for regulating the temperature of planets in response to variations in solar brightness i.e.
variation in the distance of the planets from the sun in response to changes in residual gravitation arising from
changes in temperature.

At the same time this state of affairs obviates the logical necessity of space travel undertaken to get away from
dying stars in search of other realms near younger stars. This finally leads to a resolution of the Fermi paradox
\cite{Baugher} ''If intelligent life can evolve so easily in the universe then where is the evidence of its visit to 
the earth?". 

In view of the above discussion supporting the possibility for long term survival of life on earth having been well
provided for by the Creator, the argument of Dyson \cite{Dyson} for realization of space travel: ''There is 
nothing so big or so crazy that one out of million technological societies may not be driven to do so provided it is 
physically possible" becomes somewhat less compelling.

\section{Answers suggested by the Model to the Questions}

Returning to the questions posed in the first section we can now summarize the
answers suggested by the model.
There is no left right as-symmetry. The Lagrangian of the model has the bi-modal 
discrete symmetry introduced in section (\ref{parity}), page (\pageref{parity}). 
However, the operation of parity 
transformations defined there is most likely not implementable by an operator in
the (Hilbert) space of states. We are constrained to enlarge our notion of symmetry. 

There is an almost unobservable universe with a preponderance of
anti-matter having features completely symmetric to that of the presently known
particles (leptons, quarks, light quanta, massive vector bosons) which interpenetrates
our universe and yet does not produce mutual annihilation

Notice that matter of type I ( type II) and anti-matter of type II ( type I) can 
coexist the same regions of space without producing rapid  mutual 
annihilation. This is because there is a barrier against the close approach of atomic 
nuclei of type $I$  matter having $Q_1 >0,\; Q_2 > 0 \;$ and nuclei
of type II  anti-matter having $Q_1 < 0,\;Q_2 < 0 $. This barrier
is provided by the intervening shells of $e_1\;$s  ( $e_2\;$s ) having $Q_1 < 0,\; Q_2 = 0 \;( Q_1 =0 \;Q_2 > 0)$
surrounding nuclei of type I matter ( type II anti-matter) atoms. The shells of $e_1\;$s and $e_2\;$s,
however, can have no direct influence on each other since they carry different types of 
charges which are coupled to different kinds of photons. 

There are supernovas in the symmetric universe which are
not observable directly.  This is because there is another kind
of light which fails to interact with the $e_1$ s of type I matter 
which are at the foundation of all observable chemical processes and
the interdependent web of life and environment in our world \cite{Lovelock, Lovelock2}.
\\
\\
It is reasonable to hope for an understanding the apparent solar neutrino deficit \cite{Adelberger} without invoking
a mass for the neutrino for which there is no direct experimental evidence at present and which moreover introduces 
an element of arbitrariness one would prefer to avoid. The model provides us with enough flexibility to suggest 
[see, in particular, the discussion of scenarios $X_{11}, \; X_{12}$ on page (\pageref{scenx})]
that this would happen when the
following impediments to the application of the model to elementary particle physics 
data are removed.

a: Actual data from the various neutrino experiments are made freely available to all seeking
access to it. 

b: The mis-interpretation of the muon lifetime data for muon decay in optical fibres 
recently pointed out by Widom, Srivastava and Swain  \cite{Widom} is  widely noticed and
corrected. This has bearing on the values of the constants in the standard model.
 
What are the gamma ray bursts?  Could they be novae or supernovae in
the symmetric universe which look different to us because of the
second kind of light? The present model suggests the following mechanism for gamma ray bursts,
invisible dark matter in galactic halos and on intergalactic scales.

$H_4$: There are regions with Type II matter and Type II
anti-matter which intersperse regions with Type I matter and anti-matter.

Thus the distribution of total energy released in the gamma ray bursts and their frequencies 
of occurrence  simply provide information on the size distribution of matter and
antimatter fragments, their velocity and volume distributions 
in a flat non-expanding universe. The idea of a flat non-expanding
universe in the context of a scalar theory of gravity has been studied as a viable 
alternative to problematic aspects of the standard 
model associated with gravitational singularities \cite{Khan1, Khan4}. 

Parenthetical remarks: [The scalar theory of gravity \cite{Khan4} in flat
space has no singularities (unlike the Newtonian and Einsteinian theories) and reproduces the three standard results 
of Einstein's gravity \cite{Einstein}, viz. the perihelion precession of mercury, deflection of light ray due to solar
gravity and the Shapiro time delay \cite{Shapiro} of radio signals grazing the sun. The Taylor-Hulse pulsar's 
behavior \cite{TaylorH, TaylorW} is enigmatic according to this theory of gravity, however.

It is important to realize that large scale distribution of galaxies have not shown any evidence for {\em variation 
in galactic numbers} inside cosmic spherical regions centered around us {\em with the radii} of those regions. 
Such variations, had they been noticed, could have been interpreted, through the Cosmological Principle, as evidence 
for curvature in three dimensions. The large scale flatness of space is now a generally accepted, but not explicitly 
acknowledged, fact about the cosmos \cite{Guth}, \cite{Linde}, \cite{Steinhardt}. Also, it is instructive to recall that
according to the idealist Kant \cite{Kant} ,who was the first to give us the picture of the milky way as an island 
universe of stars being viewed from a location near its periphery, flat three dimensional space is an {\em a priori}
mental category for human thought processes. Indeed, without such a starting point, Kant could not have reasonably
suggested the picture of the galaxy that is now such an integral part of the universally accepted world view!
The mathematical notion of space, of course, does allow curved spaces, since Euclid's axiom of parallels and its negation 
are both compatible with
all the other axioms. But the question not addressed so far in this context is the
following. Is there another axiom apart from the axiom of parallels which has been
overlooked in the set of axioms adopted to simulate the mathematical model of physical space?  There are hints
suggesting that there is such an overlooked axiom which has the consequence that
the negation of Euclid's parallel postulate in three dimensions
is no longer compatible with the other axioms and that the duality aspects of quantum mechanics do not seem
compatible with the present model of space because of this oversight.]

Finally, we need to briefly mention the possibility that some of the
gamma ray bursts could be due to the nuclei of type II anti-matter atoms getting
annihilated in the earth's atmosphere. If so, then mobilization of a world-wide endeavor for harvesting 
these atoms as a clean source of energy is an obvious suggestion. These atoms of type II anti-matter 
could also be producing a mis-interpreted signal in the underground detectors. The
effects of the intervening shells of $e_1\;$s and $e_2\;$s on the non-corresponding nuclei
captured by them needs to be calculated before definite predictions of
characteristics of the signals can be made. This has not been completed yet.

What is the nature of CBR? 

Proposed answer: It
is due to the second kind of light interacting with the baryons and anti-baryons of the dark
matter in our vicinity  producing a gentle shaking up of the barons and anti-baryons acting as
transmitters of our kind of light.

The CBR is not due to a primordial initial singularity which, as mentioned earlier,
is not needed to explain the cosmological redshift. The microwave background is due to the
gentle shake up of atomic nuclei which are immersed in the ambient light of the second kind 
emanating from the (locally dominant) interpenetrating fragment of cosmos of type II matter and anti-matter in our
vicinity. See also the discussion of the Great Attractor problem in section (\ref{sec-8.2}), page (\pageref{sec-8.2}) 
which corroborates this viewpoint and explains the near equality of magnitudes and opposite directions for
velocities of the sources of CBR ( predominantly of type II matter and antimatter) and of the local 
distribution of (type I) matter.

\section{Proposed Experiments and Related Topics}

\subsection{Storage Ring Experiment}

Let us assume first that matter on earth and the solar system is of type III.
We could then try to confirm this by doing the experiment described below.
We could take protons and nuclei out of atoms in such a way that not only
$e_1\;$s but also $\overline{e}_2\;$s are shaken out of the atoms.  At
present, we are making protons and nuclei using our electromagnetic
technology, which is based on $e_1\;$s and the first kind of light.
Our consciousness is locked into the collective behavior of the
$e_1$'s in our bodies.  So this will not help us get rid of the
$\overline{e}_2$'s clinging on to the protons and nuclei made in
various devices based on electromagnetism associated with $e_1\;$s and
the first kind of light---devices which are manipulated by hands and
eyes locked into consciousness  (or subjective attention) attuned to $e_1\;$s and
the first kind of light.  Although our devices and bodies and
everything on earth, sun and stars are flooded with $\overline{e}_2\;$s
clinging to the nuclei (if they are made of type III matter), we have been 
unable to detect these so far - but only until now, since it is possible to detect
them by initiating the following sequence of manipulations using our existing
capabilities.

We take the nuclei and using existing technology, strip them of $e_1\;$s
first, and then accelerate them with the the same technology for
storage in rings containing these nuclei moving in
counter-rotating directions.  This is the first stage.  In the second
stage, the nuclei moving in opposite directions are arranged to
collide with each other, so that the second kind of electromagnetism
between $\overline{e}_2\;$s and nuclei is brought into play.
Recall that $\overline{e}_2$'s have $Q_2$ charge of $-1$ and
nuclei have $Q_2 > 0$.  This would remove the
clinging $\overline{e}_2$ and keep the nuclei intact---provided
the collisions are arranged to be not too violent.

Now these nuclei, which have been completely or partially
stripped of $\overline{e}_2\;$s could be stored in one of the rings
containing these protons/nuclei. It can be arranged that they maintain constant
speed by providing electrical energy from our present devices
(i.e. the microwave cavities around storage rings of our present day
accelerators).  Nuclei are being accelerated and produce
electromagnetic radiation of both kinds since $Q_1> 0 \;  Q_2  > 0$ for these
the nuclei.  We can now measure
the rate of energy transfer to the nuclei to compensate for the energy loss due
to radiation from accelerated nuclei through emission of photons of both kinds.   This rate of
energy input into nuclei to maintain their constant speed can presumably be
measured easily by reading the power meters provided by the engineers. Thus for nuclei this 
rate of energy consumption in the storage ring would be expected to be more than what
it would be  if $Q_2=0$ by a factor $\frac{Q_1^2 + Q_2^2}{Q_1^2}$.

If matter on earth is of type I then we need only strip the nuclei of 
$e_1\;$s using our electromagnetic technology. There are no ${\overline{e}}_2\;$s
attached to the nuclei and we can skip the second stage and proceed directly to 
the last stage in the  experiment outlined above.

\subsection{Great Attractor}
\label{sec-8.2}

As a preliminary to a description of the second experiment, we give the following quotation
from Peebles \cite{Peebles} which describes the so called 'Great Attractor' in astrophysics and describe
an alternative proposed explanation for it.
\begin{quote}
``The motion of the Solar system relative to the frame in which CBR is
isotropic is $v_\odot - v_{CBR} = 370 \pm 10$ km/s, to $\alpha = 11.2$
h, $\delta = -7^\circ$, $\ell = 264.7 \pm 0.8^\circ$, $b = 48.2 \pm
0.5^\circ$.  The conventional correction for the solar motion relative
to the Local Group is $300$ km/s to $\ell = 90^\circ$, $b=0$.  This is
close to the mean motion defined by the Local Group and to the
velocity that minimizes the scatter in the local distance-redshift
relation.  With this correction the velocity of the Local Group relative
to CBR is $600$ km/s toward $\alpha = 10.5$ h, $\delta = -26^\circ$,
($\ell = 268^\circ$, $b=27^\circ$).  This velocity is much larger than
the scatter in local redshift-distance relation" 
\end{quote}

We need not invoke a fictitious Great Attractor to explain why the motion of
the source generating CBR is with a velocity equal in magnitude and opposite in 
direction to that of the matter in the local group. Besides, the Great Attractor 
hypothesis has no explanation for the equality of the two velocity magnitudes.
A more plausible explanation is as follows. 

There is  a vast region of dark matter atoms (assumed to be composed of either $\overline{e}_2$
and baryons or $e_2$ and anti-baryons, i.e. type II matter or anti-matter) in juxtaposition with 
ordinary matter atoms (assumed to be of type I matter and composed of $e_1$ and baryons) constituting
the visible objects of the Local Group. The ordinary matter is partaking of the overall (large scale) 
motion of dark matter moving around it. If it is now assumed that the situation is analogous to that
in which it seems ''as if" (*) the sense of time flow is
reversed for dark matter relative to ordinary matter (type $I$) then it would appear to be moving
in a direction opposite to that of the surrounding ordinary matter with a velocity equal in magnitude
to that of the ordinary matter. In this picture the source of CBR would then be the dominant dark
matter component of our cosmic neighborhood. More explicitly, the CBR effect observed is due to the
interaction of second kind of light (${\cal{A}}_\mu^2$) produced by dark matter atoms with the
baryons $(Q_1 > 0, Q_2 > 0)$ or antibaryons  $( Q_1 < 0, Q_2 < 0)$ in the predominantly type II matter
or anti-matter in our cosmic neighborhood.

(*) An explanation of what is really meant by  the above statement "analogous to that in which it seems ''as if" 
the sense of time flow is reversed .." will now be given. Let us first emphasize that nothing 
mysterious is being suggested by this statement. All that is being said is that in statistical
mechanical systems, under appropriate boundary conditions, it may happen that the motion of a sub-system
acquires characteristics that one might usually associate with the backward evolution in time of  
the system. Thus for example, in most laboratory situations waves propagate outwards from 
a point. However, one can easily subject matter to unusual boundary conditions in which its behavior has the
appearance produced by a cylindrical wave propagating from the boundary of the cylindrical
container to its axis. The reader might try observing water in a pot placed on an old 
washing machine which vibrates with a large amplitude when it is run.

\subsection{Superconductivity Experiment}
\label{supercond}

In preparation for the description of our second experiment, we next ask

Question: Are the Cooper pairs in superconductor really consisting of two electrons with oppositely directed
velocities of equal magnitude?

Proposed Answer: No. They are composed of $e_1$ and $\overline{e}_2$ (present in matter of type III or possibly
type I at low temperatures) undergoing a similar overall motion inside a superconductor.  The reason their
velocities seem oppositely directed is because the $\overline{e}_2$'s are moving with a sense of time flow reversed 
relative to the $e_1$'s (analogous to the phenomena associated with the Great Attractor described above). 
Thus the interaction between the pairs ($e_1, \overline{e}_2$), locked in oppositely directed motions, occurs here also 
through the intermediacy of nuclei, just as it does in the BCS theory of superconductivity \cite{Bardeen}. However
the actual mechanism is different. It is not that of phonons associated with nuclei. Rather it is due to both kinds 
of charges carried by the nuclei.

Notice that this proposed mechanism can explain in a very natural fashion the experimentally observed separation
of regions of charge and spin localisation in high temperature superconductors \cite{Moritomo, Tranquada1, Tranquada2}
subjected to neutron diffraction analysis. There is no great difficulty in comprehending this phenomena within the 
framework of a two component model involving $e_1$'s and $\overline{e}_2$'s, in contrast to the usual models based
on $e_1$'s only which seem too artificial. 

We now describe the following experiment that may be set up to test for this proposed answer which is suggested by
the model of elementary particle interactions presented in this paper. 

We  take a long superconductor. We stir up $e_1$'s at one end of it using electromagnetic fields
associated with light of the first kind. When $e_1$'s are accelerated by the applied electromagnetic fields they
produce an overall motion of $\overline{e}_2$, occurring
through the intermediacy of nuclei with both kinds of charges.
The accelerated $\overline{e}_2$ are then expected to emit light of the second kind, which
can be reflected by a mirror made of a good reflector for ${\cal{A}}_\mu^2$ to a remote replica of the
superconductor.  Notice that a chemically homogeneous nuclide for ${\cal{A}}_\mu^1$ is not necessarily chemically
homogeneous for ${\cal{A}}_\mu^2$ [ as explained in section (\ref{nuclei}), page (\pageref{nuclei})] and the search 
for a good mirror for ${\cal{A}}_\mu^2$ would involve a bit of trial and error. Once this has been successfully achieved
one expects to see flashes of ordinary light $({\cal{A}}_\mu^1)$ from the remote superconductor as the 
shaking up of $e_2$'s by the incident ${\cal{A}}_\mu^2$ induces motion of $e_1$'s through the intermediacy
of nuclei carrying both kinds of charge.
\\
\\

\section{Non-coding Segments of the DNA}
\label{DNA}

If matter on earth and the solar system is of type III (including living 
matter on earth) then, as explained above, 
the usual electromagnetic fields are being generated with devices assembled
by us with eyes sensitive to
${\cal{A}}_\mu^1$, because our consciousness must have been locked to
chemical processes in the retina produced by $e_1$'s interacting with
${\cal{A}}_\mu^1$.  Chemical processes  produced by
$\overline{e}_2$'s interacting with ${\cal{A}}_\mu^2$ are also 
present (in matter of type III and presumably also in matter of type I at low temperatures).
However the effect of the latter chemical processes on consciousness would be indirect (i.e. through the
intermediacy of interactions of ${\cal{A}}_\mu^2$ with nuclei carrying both kinds of charge).Thus it is understood that
although the chemical processes involving $e_1$'s and $\overline{e}_2$'s may be similar in intensity and frequency of occurrence, the effects on
consciousness may be quite different in their vividness. Since the chemical processes involving $\overline{e_2}$ 
and ${\cal{A}}_\mu^2$ 
influence consciousness indirectly
unlike the direct effect on consciousness of processes involving $e_1$'s and ${\cal{A}}_\mu^1$, the question of 
a possible mechanism for this is now raised. It is suggested that this could have
been arranged by Nature through a segregation of the segments of DNA specialized in exercising chemical
control via one or the other of the two parallel chemical mechanisms. If this is so then it
would  explain why some segments of the DNA \cite{DNA} can not be analyzed by the usual methods of
chemical analysis (attuned to human consciousness) to determine their function and they would appear to be non-coding 
and dormant.

Finally, let us emphasize that the nature of physical reality may well be more complex than a simple 
once and for all selection between the various scenarios. Thus, for example, non-living matter in  
the solar system (except the earth)  may be type I. However living matter on earth may be type III.
\\
\section{Charged Neutrino}
\label{neutrino}

The model has one neutrino with $Q_1 = 1 $ and $Q_2 = 1$ associated with each family. 
Have these neutrinos been seen?
We propose that the jets of hadrons produced in $e_1, \overline{e}_1$
collisions are not due to constituent quarks since the masses of the hypothetical quarks turn out to be
quite small compared to the hadron masses. If quarks really have the low masses suggested by the jet 
phenomenology and they are almost non-interacting at short range \cite{coleman2} then the large masses of hadrons
(presumed to be lowest energy states of quark systems) are quite problematic. We believe that we can not 
avoid this paradox by  introducing the hypothetical
mechanism of quark confinement which has not received explicit demonstration so far. The so called
mass gap \cite{Jaffe} problem in which mathematical physicists have been interested (and whose precise formulation is
as of now not available), is most likely a reformulation of this paradox.

So if the jets can not be due to quarks then they may be something else. The most natural candidates
suggested by this model are the charged neutrinos. In fact we believe that the study of jets may be 
used to understand the quantum field theory of massless charged particles. \cite{Lee, SW}

\section{Criticism of the Model and Conclusion}

In conformity with the prevalent ideas we  still have only the weak requirement of 
''renormalizability" \cite{Schweber, Bollini, Ashmore, Zuber} for the model.
However in agreement with Dirac's views \cite{Dirac}, we believe that
renormalization in a fundamental theory
is to be regarded as a 
calculable and physically measurable departure from a conceptually simpler 
situation - in other words renormalization effects must all be finite.
In the present model, although some of the infinities arising in renormalization integrals
(other than those associated with the anomalous Feynman diagrams \cite{Bouchiat, GrossJ, GG}) do cancel, other
infinities survive. We believe that the model must eventually evolve to the stage in which all divergences
of Feynman diagrams cancel (not just those of the anomalous diagrams).  This may be possible,
since the model provides hints of new kinds of symmetries linking
scalars, fermions, and vector bosons (these symmetries are implemented differently from supersymmetry \cite{Wess}).

Dirac's dream of a finite quantum field theory is closer to realization.

Failure to realize this dream has led some of us  to propose 
the radical viewpoint that there are various levels of descriptions with their own
laws. However we feel that if the non existence of a calculable mechanism to derive laws at any stage from
those at a more fundamental level is accepted as a matter of principle, then what we
are really saying is that there is no physical theory and
that it is just an exercise in very complicated parameter fitting of
experimental data.
\\
\\
\\
\\
\\
\\
\\
\newpage

\section{Appendix}
\label{appendix}

Representation for the Dirac $\gamma$-matrices are taken to be

\begin{eqnarray}
\begin{array}{c}
\gamma^0  = 
\left\lgroup
\begin{array}{cc}
0 & I\\
I & 0
\end{array}
\right\rgroup \;\gamma^i  = 
\left\lgroup
\begin{array}{cc}
0 & {\sigma}^i\\
-{\sigma}^i & 0
\end{array}
\right\rgroup\\[0.1 in]
\gamma_5=i\gamma^0 \gamma^1 \gamma^2 \gamma^3  = 
\left\lgroup
\begin{array}{cc}
-I & 0\\
0 & I
\end{array}
\right\rgroup
\end{array}
\end{eqnarray}

so that

\begin{eqnarray}
&{\gamma^0}^\dag =  \gamma^0,\; \;{\gamma^i}^\dag = -\gamma^i&\\[0.1 in]
&{\gamma^0}^2 =  -{\gamma^i}^2 =  \left\lgroup
\begin{array}{cc}
I & 0\\
0 & I
\end{array}
\right\rgroup = \cal{I}&
\end{eqnarray}

In this representation the matrix C defined by

$
C {{\gamma}^{\mu}}^* C^{-1} = {-\gamma}^{\mu}
$

and

$
C^\dag= -C
$

is given by  $C= \pm \gamma^2$. 

Thus $C^* = - C,\;\;\; C^2 = -{\cal I},\;\; \; C^t = C$

[ $*$ is complex conjugation, $t$ is transposition and $\dag$ is
hermitian conjugation of a matrix]

This matrix  C has the property that if q transforms under
Lorentz transformations as a Dirac spinor then $C q^*$ also transforms as a Dirac spinor.

The left handed component of $q$ is defined as

$
q_L= \frac{1}{2}(1 + \gamma_5) q
$

The right handed component of $q$ is defined as

$
q_R = \frac{1}{2}(1 - \gamma_5) q
$

We define ${q'}_R, {q'}_L$ as

$
{q'}_R = C {q_L}^*, \;  \; {q'}_L = C {q_R}^*
$

Then

$
q_L=C {{q'}_R}^*,\; \; q_R=C{{q'}_L}^*
$

The matrix C defined above also
satisfies the relation

$
C \gamma^0 {\gamma}^{\mu} C^{-1} = (\gamma^0 {\gamma}^\mu)^t
$

The above relations imply (for anti-commuting spinors q, r) that 

\begin{eqnarray}
\begin{array}{rcl}
\label{appendeq}
\overline{q}_R {\gamma}^{\mu} r'_R & = & - \overline{r}_L {\gamma}^{\mu} {q'}_L\\[0.1 in]
\overline{q'}_R {\gamma}^{\mu} r'_R & = & - r_L {\gamma}^{\mu} q_L\\[0.1 in]
\overline{q}_R {\gamma}^{\mu} r_L & = & \overline{q'}_R {\gamma}^{\mu} r'_L\; = \; 0\\[0.1 in]
\overline{q}_R {\gamma}^{\mu} r'_L & = & \overline{q'}_R {\gamma}^{\mu} r_L\; = \; 0

\end{array}
\end{eqnarray}

The actions of generators of $R_1$ and $R_2$ on the multiplet ($q_h,r_h,{q'}_{\tilde{h}}, {r'}_{\tilde{h}},$${s'}_{\tilde{h}},
{t'}_{\tilde{h}}, s_h, t_h)$ are deducible from (\ref{variation4}).

For the generators $T^i(S^i)$ of $R_1(R_2)$ the actions on the multiplet , expressed as a column vector, are given by 
multiplication of the column vector on its left with the matrices $F^i(G^i)$ which are as follows:
[We have set the values of the parameters $\xi_1, \xi_2, \eta_1, \eta_2$ to +1, which have turned out to
be the only values of relevance at this stage]

\begin{eqnarray}
F^i & = &
\frac{1}{2}
\left\lgroup
\begin{array}{cccc}
{\sigma}^i & 0 & 0 & 0\\
0 & -{{\sigma}^i }^*& 0 & 0\\
0 & 0 & {\sigma}^i & 0\\
0 & 0 & 0 & -{{\sigma}^i}^*
\end{array}
\right\rgroup
\end{eqnarray}

\begin{eqnarray}
\begin{array}{rcl}
&G^3  = 
\frac{1}{2}
\left\lgroup
\begin{array}{cccc}
I & 0 & 0 & 0\\
0 & -I & 0 & 0\\
0 & 0 & I & 0\\
0 & 0 & 0 & -I
\end{array}
\right\rgroup&\\[0.15 in]
&G^1  = 
\frac{1}{2}
\left\lgroup
\begin{array}{cccc}
0 & 0 & 0 & {\sigma}^2\\
0 & 0 & {\sigma}^2 & 0\\
0 & {\sigma}^2 & 0 & 0\\
{\sigma}^2 & 0 & 0 & 0
\end{array}
\right\rgroup&\\[0.15 in]
&G^2  = 
\frac{-i}{2}
\left\lgroup
\begin{array}{cccc}
0 & 0 & 0 & {\sigma}^2\\
0 & 0 & -{\sigma}^2 & 0\\
0 & {\sigma}^2 & 0 & 0\\
-{\sigma}^2 & 0 & 0 & 0
\end{array}
\right\rgroup&
\end{array}
\end{eqnarray}

Using the relation 
$
\sigma^2 {\sigma^i}^* = - \sigma^i \sigma^2
$

one may verify that these matrices satisfy the commutation relations

\begin{eqnarray}
\begin{array}{rcl}
\left[F^i, F^j]\right] & = & i \epsilon^{ijk} F^k\\
\left[G^i, G^j \right] & = & i \epsilon^{ijk} G^k \\
\left[ F^i, G^j \right] & = & 0
\end{array}
\end{eqnarray}

\section{Acknowledgements}

I would like to thank the College of Bahamas for the opportunity to teach physics at the College.
I would also like to express my deep appreciation to Fayyazuddin, Karen Fink, Yogi Srivastava, Mahjoub Taha, 
Allan Widom, Husseyin Yilmaz, Abdel-Malik AbderRahman, Mohammed Ahmed, Asghar Qadir, Qaisar Shafi, Abner Shimony,
Badri Aghassi, Peter Higgs, Derek Lawden, Robert Carey,  Priscilla Cushman, Stephen Reucroft, 
Tahira Nisar, John Swain, Alan Guth, Fritz Rohrlich, Martinus Veltman, Kalyan Mahanthappa, Claudio Rebbi, 
Stephen Maxwell, Sidney Coleman, Christian Fronsdal, Moshe Flato, John Strathdee, Muneer Rashid, Klaus Buchner, Marita Krivda,
Henrik Bohr, Lochlaimm O'Raifeartaigh, Uhlrich Niederer, Werner Israel, Patricia Rorick, Yasushi Takahashi, Helmut
Effinger, John Synge, John Wheeler, Nandor Balazs and the {\em departed souls: Abdus Salam, Marvin Friedman,
Cornelius Lanczos, Asim Barut, Jill Mason,
Sarwar Razmi, Nicholas Kemmer, Paul Dirac, Peter and Ralph Lapwood, Iqbal Ahmad, Naseer Ahmad} (may they be favored by Allah) for friendship 
and/or moral support and/or conversations on various occasions. 

I am grateful to my son Bilal for his cheerful involvement in my life, advice and help during the preparation of the 
manuscript.

\newpage
\bibliography{nucleus1}

\end{document}